\documentclass[twocolumn,preprintnumbers,aps,prd,superscriptaddress,footinbib,tightenlines, hidelinks]{revtex4-1}

\usepackage{amsmath}
\usepackage{amsfonts, amsthm, amssymb, graphicx, tikz}

\usepackage{xcolor}
\usepackage[utf8]{inputenc}
\usepackage[normalem]{ulem} 

\usepackage{wrapfig}
\usepackage{mathrsfs}
\usepackage{setspace}
\usepackage{physics}
\usepackage{hyperref}
\hypersetup{
    colorlinks=false,
    linkcolor=black,
    linktoc=all
}

\usepackage{placeins}

\usepackage{bbm}

\usepackage[framemethod=TikZ]{mdframed}
\mdfdefinestyle{boxed}{%
    linecolor=black,
    outerlinewidth=0.5pt,
    roundcorner=5pt,
    innertopmargin=\baselineskip,
    innerbottommargin=\baselineskip,
    innerrightmargin=20pt,
    innerleftmargin=20pt,
    backgroundcolor=gray!05!white
    }
    
\newcommand{\A}{\mathcal{A}}

\newcommand{\bfp}{\mathbf{p}}

\renewcommand{\d}{\,\mathrm{d}}

\newcommand{\nn}{\nonumber}

\definecolor{light-gray}{gray}{0.95}

\def\be{\begin{equation}}
\def\ee{\end{equation}}
\def\ba{\begin{eqnarray}}
\def\ea{\end{eqnarray}}

\def\bq{\begin{quote}}
\def\eq{\end{quote}}

 at 10truept
\newcommand{\bfx}{{\bf x}}
\newcommand{\bfy}{{\bf y}}

\newcommand{\bfk}{{\bf k}}

\newcommand{\beq}{\begin{equation}}
\newcommand{\eeq}{\end{equation}}
\newcommand{\bea}{\begin{eqnarray}}
\newcommand{\eea}{\end{eqnarray}}
\newcommand{\beqa}{\begin{eqnarray}}
\newcommand{\eeqa}{\end{eqnarray}}

\def\d{{\rm d}}

\usetikzlibrary{patterns}
\usetikzlibrary{shapes}
\usetikzlibrary{arrows,shapes,positioning}
\usetikzlibrary{decorations.pathmorphing}
\usetikzlibrary{decorations.markings}
\tikzset{cross/.style={cross out, draw=black, minimum size=2*(#1-\pgflinewidth), inner sep=0pt, outer sep=0pt},
cross/.default={1pt}}

\tikzset{zigzag/.style={decorate, decoration=zigzag}}

\tikzstyle arrowstyle=[scale=1]
\tikzstyle directed=[postaction={decorate,decoration={markings,
    mark=at position .65 with {\arrow[arrowstyle]{stealth}}}}]
\tikzstyle reverse directed=[postaction={decorate,decoration={markings,
mark=at position .65 with {\arrowreversed[arrowstyle]{stealth};}}}]

\begin{document}
\raggedbottom

\title{
Positivity Bounds without Boosts
}

\author{Tanguy Grall} 
\affiliation{DAMTP, Centre for Mathematical Sciences, University of Cambridge, CB3 0WA, United Kingdom}
\author{Scott Melville} 
\affiliation{DAMTP, Centre for Mathematical Sciences, University of Cambridge, CB3 0WA, United Kingdom}

\date{\today}

\begin{abstract}
\noindent
We derive the first positivity bounds for low-energy Effective Field Theories (EFTs) that are not invariant under Lorentz boosts.
``Positivity bounds'' are the low-energy manifestation of certain fundamental properties in the UV---to date they have been used to constrain a wide variety of EFTs, however since all of the existing bounds require Lorentz invariance they are not directly applicable when this symmetry is broken, such as for most cosmological and condensed matter systems. 
From the UV axioms of unitarity, causality and locality, we derive an infinite family of bounds which (derivatives of) the $2\to2$ EFT scattering amplitude must satisfy even when Lorentz boosts are broken (either spontaneously or explicitly).
We apply these bounds to the leading-order EFT of both a superfluid and the scalar fluctuations produced during inflation, comparing in the latter case with the current observational constraints on primordial non-Gaussianity. 
\end{abstract}


\maketitle






\noindent From the early ages of our universe to present day condensed matter physics, the natural world favours solutions to its fundamental laws which break Lorentz invariance. 
For instance, both the inflationary spacetime background \cite{Cheung:2007st,Senatore:2010wk,Piazza:2013coa,Finelli:2018upr} and many-body systems at high density \cite{Son:2002zn,Son:2005ak,Dubovsky:2011sk,Dubovsky:2011sj,Nicolis:2011cs,Nicolis:2013lma,Delacretaz:2014jka} effectively provide a preferred reference frame, 
and low-energy fluctuations in these systems are accurately captured by Effective Field Theories (EFTs) which are not invariant under (linearly realised) Lorentz boosts.
The EFT framework has proven remarkably successful, even in systems with broken boosts: by providing a simple parametrisation of the possible low-energy interactions (in which coupling constants are fixed by comparing with data), it has allowed us to make great progress in analysing our low-energy observations even in cases where we do not fully understand the underlying UV physics (such as inflation).
 
However, care must be taken when adopting an EFT parametrisation, since not all of the EFT parameter space is equally viable. 
Over the past two decades, a powerful connection between low-energy Lorentz-invariant EFTs and their underlying high-energy UV completions, known as \emph{positivity bounds}, have been developed \cite{Adams:2006sv,Jenkins:2006ia, Bellazzini:2016xrt,deRham:2017avq,deRham:2017zjm,Remmen:2020uze,Bellazzini:2020cot,Tolley:2020gtv,Caron-Huot:2020cmc, Li:2021cjv, Arkani-Hamed:2020blm}. 
Using the causal properties of the $2\to2$ scattering amplitude to bridge between the EFT coupling constants and the unitarity of the underlying UV completion, these bounds constrain the sign or relative size of the EFT interactions. 
When these positivity bounds are violated, it signals that no new high energy physics, no matter how ingenious our model-building, can restore causality, unitarity and locality beyond the EFT cutoff. 
These UV/IR connections have been applied to great effect in a huge variety of different settings: including constraining new physics beyond the 
Standard Model~\cite{Bellazzini:2018paj,Zhang:2018shp,Bi:2019phv, Zhou:2020ubs, Yamashita:2020gtt, Fuks:2020ujk, Zhang:2020jyn, Bi:2019phv, Zhang:2018shp, Remmen:2020vts, Remmen:2019cyz}; 
chiral perturbation theory~\cite{Pham:1985cr,Ananthanarayan:1994hf,Pennington:1994kc,Distler:2006if,Vecchi:2007na,Wang:2020jxr}; 
corrections to general relativity~\cite{Bellazzini:2015cra,Cheung:2016wjt,Camanho:2014apa,Gruzinov:2006ie, Herrero-Valea:2020wxz, Alberte:2020jsk, Alberte:2020bdz}; 
massive gravity~\cite{Cheung:2016yqr,Bonifacio:2016wcb,Bellazzini:2017fep,deRham:2017xox,deRham:2018qqo,Alberte:2019xfh, Alberte:2019zhd, Wang:2020xlt}; 
higher-spin theories~\cite{Bonifacio:2018vzv,Bellazzini:2019bzh, Melville:2019tdc}; 
various scalar field theories~\cite{Nicolis:2009qm,Komargodski:2011vj,Elvang:2012st,deRham:2017imi,Chandrasekaran:2018qmx}; 
Einstein--Maxwell theory and the Weak Gravity Conjecture~\cite{Cheung:2014ega,Cheung:2018cwt,Hamada:2018dde,Cheung:2019cwi,Bellazzini:2019xts,Charles:2019qqt}.
However, one limitation of these bounds is that they rely on Lorentz invariance at all energy scales---for all their successes, they cannot be used to constrain everyday systems in which Lorentz invariance is broken. 
This is certainly not for want of trying: previous works have applied Lorentz-invariant positivity arguments to cosmology by requiring either additional assumptions about the UV \cite{Baumann:2015nta, Ye:2019oxx, Kim:2019wjo, Herrero-Valea:2019hde} or starting from a particular covariant theory which has both flat and symmetry-breaking vacua \cite{Melville:2019wyy, Kennedy:2020ehn}.
 
In this letter we derive the first positivity bounds which can be applied directly to EFTs in which Lorentz invariance is broken (either spontaneously or explicitly).
We assume only that the underlying UV completion respects: \\[5pt]
(a) the unbroken symmetries (rotations / translations), \\
(b) crossing (different scattering channels are related), \\
(c) unitarity (the optical theorem), \\
(d) analyticity (causality), \\
(e) polynomial boundedness (locality). \\[5pt]
Properties (a) and (b) fix the kinematics of the problem, and property (c) has recently been developed in theories without boosts in \cite{Grall:2020tqc} (see also \cite{Pajer:2020wnj, Stefanyszyn:2020kay, Kim:2021pbr}, as well as earlier work by \cite{Baumann:2011su, Baumann:2014cja, Koehn:2015vvy, deRham:2017aoj}). 
In this work, we tackle properties (d) and (e), and show for the first time that it is possible to define a dispersion relation for an amplitude without boosts which retains manifest crossing symmetry. 
This allows us to identify an infinite number of inequalities which constrain (derivatives of) the $2\to2$ EFT scattering amplitude.
As a practical application, we apply the simplest of these bounds to the EFT of a superfluid and to the EFT of inflation and discuss their implications for primordial non-Gaussianities.


\section{Properties of the Amplitude}
\label{sec:properties}

\noindent Let us begin by listing the properties of $\mathcal{A}_{ \pi_1 \pi_2 \to \pi_3 \pi_4 }$, the $2\to 2$ scattering amplitude between identical scalar fluctuations $\pi$, that we will use to derive positivity bounds.

~\\
{\bf (a) Spacetime Symmetry.}
We will assume that the Poincar\'{e} invariance of the system is broken by a single preferred (time-like) direction $n^\mu$. This breaks enough symmetry (i.e. Lorentz boosts) to capture condensed matter systems such as fluids and cosmological systems such as inflation, but preserves enough symmetry (i.e. spatial rotations and translations in both space and time) to retain a well-defined $S$-matrix description of the scattering between asymptotic states. 
We will additionally assume that, in the absence of interactions, the equation of motion for the scalar $\pi$ at low energies takes the form,
\begin{align}
 \omega^2  = c_\pi^2 |\bfk|^2 + m^2   \; , 
 \label{eqn:free_propagation}
\end{align}
where $c_\pi$ is a fixed constant speed (which may differ from the invariant speed $c$ of the broken Lorentz boosts), $m$ is an invariant mass, and we are working in momentum space so that $i \partial_\mu$ acting on $\pi$ becomes $k_\mu = (\omega, \bfk)$, where $\omega = n^\mu k_\mu$ is the time-like component (i.e. frequency) and $\bfk$ are the remaining spatial components (i.e. wavenumbers).
The asymptotic states for our scattering process are therefore labelled by their spatial momentum, $\bfk_j$, with their frequencies $\omega_j$ fixed by \eqref{eqn:free_propagation}, where we adopt a convention in which $\omega > 0$ ($<0$) for incoming (outgoing) fluctuations. 
The low-energy effective interactions are built from both invariant contractions $\partial_\mu \partial^\mu$ as well as time-like derivatives $n^\mu \partial_\mu$ acting on $\pi$, and a natural basis of kinematic variables is therefore $s_{ij} =   (\omega_i + \omega_j)^2 - c_\pi^2 | \bfk_i + \bfk_j |^2$ and $\omega_{ij} = \omega_i + \omega_j$. These are not all independent, since e.g. time translations sets $\omega_{12} = \omega_{34}$ and similarly spatial translations set $s_{12} = s_{34}$. A complete basis is given by the five variables
\footnote{
Note that we have defined $s$ and $t$ using the effective metric $Z^{\mu\nu} = \text{diag} (-1, c_\pi^2, c_\pi^2, c_\pi^2 )$ which determines the free propagation of $\pi$ in equation \eqref{eqn:free_propagation}, rather than the Minkowski metric associated with the broken boosts, which would be $c_s^2 \tilde{s}  = s - (1 - c_s^2 ) \omega_s^2$ and $c_s^2 \tilde{t}  = t - (1 - c_s^2 ) \omega_t^2$, where $c_s = c_\pi / c$ is the dimensionless ratio of the low-energy sound speed to the invariant UV sound speed. 
Our assumption that the dispersion relation for $\pi$ (i.e. the free theory) has an accidental Lorentz invariance (transformations which preserve $Z^{\mu\nu}$) is what allows the use of relativistic Mandelstam-like variables---in cases where the dispersion relation is far from linear, there would be little motivation for forming $s$ and $t$ combinations and in that case the non-relativistic problem is best treated in terms of the energies directly (c.f. the Kramers-Kronig relations). 
},
\begin{align}
s = s_{12}  ~ , ~  t = s_{13}  ~, ~  \omega_s = \omega_{12}  ~, ~ \omega_t = \omega_{13} ~, ~ \omega_u = \omega_{14} 
\end{align}
where $u = s_{14}$ is fixed by $s +t + u = 4 m^2$.  
For the interaction, $\pi_1 \pi_2 \to \pi_3 \pi_4$, which we refer to as the \emph{$s$-channel process} with amplitude $\mathcal{A}_{\pi_1 \pi_2 \to \pi_3 \pi_4} =: \mathcal{A}_s (s,t,\omega_s, \omega_t, \omega_u)$, real physical momenta correspond to the domain $s > -t > 0$ and $\omega_s > \sqrt{ s + ( \omega_u - \omega_t )^2 }$, which we refer to as the \emph{$s$-channel region}.


~\\
{\bf (b) Crossing.}
In addition to the $s$-channel region, there are five other regions in which all of the momenta can become real. These correspond to the $2 \to 2$ processes,
\begin{align}
&s: \; \pi_1 \pi_2 \to \pi_3 \pi_4  ,
 &&t: \; \pi_1 \bar{\pi}_3 \to \bar{\pi}_2 \pi_4,
&&u: \; \pi_1 \bar{\pi}_4 \to \pi_3 \bar{\pi}_2,  \nonumber \\
&\bar{s} : \; \bar{\pi}_3 \bar{\pi}_4 \to \bar{\pi}_1 \bar{\pi}_2 ,
&&\bar{t}:\; \bar{\pi}_4 \pi_2 \to \pi_3 \bar{\pi}_1,
&&\bar{u} : \; \bar{\pi}_3 \pi_2 \to \bar{\pi}_1 \pi_4  , \nonumber
\end{align}
where $\bar{\pi}$ is the charge-conjugate field ($=\pi$ for a real scalar). 
For example, the $u$-channel process corresponds to the $u$-channel region $u > -t > 0$ and $\omega_u > \sqrt{u + (\omega_s - \omega_t )^2}$, and the associated amplitude is $\mathcal{A}_u (u,t, \omega_u, \omega_t, \omega_s)$. 
Since we are considering the scattering of identical scalar fluctuations, all six channels are physically equivalent. 
This leads to relations between the amplitude at different kinematics, e.g. \footnote{
Note that in the Lorentz-invariant case it is not necessary to distinguish between $u$ and $\bar{u}$, so crossing $\pi_1 \leftrightarrow \pi_3$ or $\pi_2 \leftrightarrow \pi_4$ lead to the same relation for the amplitude. In contrast, \eqref{eqn:crossing} corresponds to crossing $\pi_2 \leftrightarrow \pi_4$, and had we instead performed $\pi_1 \leftrightarrow \pi_3$ we would have found, $\mathcal{A}_s (s,t,\omega_s, \omega_t,\omega_u) = \mathcal{A}_{\bar{u}} (u,t, -\omega_u, -\omega_t, -\omega_s)$, which is related to $\mathcal{A}_u$ by a time reversal. 
},
\begin{align}
   	\A_s (s,t,\omega_s,\omega_t,\omega_u)&=\A_{u} (u,t, \omega_u, \omega_t, \omega_s) \; .
   	\label{eqn:crossing}
\end{align} 
 
~\\
{\bf (c) Unitarity.}
Unitarity of the $S$-matrix leads to the well-known optical theorem for scattering amplitudes, 
\begin{align}
2 \text{Disc} \, \mathcal{A}_s = \sum_n \mathcal{A}_{\pi_1 \pi_2 \to n} \mathcal{A}^*_{\pi_3 \pi_4 \to n}
\label{eqn:optical}
\end{align}
which relates the discontinuity,
$ \text{Disc} \, \mathcal{A}_s  = \tfrac{1}{2i} \left(  \mathcal{A}_s  - \mathcal{A}_{\bar{s}}^* \right) $,
to a sum over amplitudes for the inelastic processes $2 \to n$.
We will require that the underlying high-energy physics which UV completes the EFT is unitarity---in particular, the optical theorem then guarantees that $\text{Disc} \, \mathcal{A}_s  \geq 0$ for forward-limit scattering (i.e. when $\pi_1 \pi_2$ coincides with $\pi_3 \pi_4$) in the UV.

~\\
{\bf (d) Analyticity.}
\emph{Causal} interactions correspond to \emph{analytic} response functions \cite{Eden, Camanho:2014apa}. 
For Lorentz-invariant scattering amplitudes, causality translates into the condition that $\mathcal{A}_s (s,t)$ is analytic in the complex $s$-plane at fixed $t$ (for all $\text{Im} \, s \neq 0$). 
This is a key ingredient in deriving Lorentz-invariant positivity bounds, used to relate the EFT amplitude (at small $s$) to the underlying UV amplitude (at large $s$) via Cauchy's residue theorem.

However, when boosts are broken the amplitude is also a function of $\{ \omega_s$, $\omega_t$, $\omega_u\}$ 
and one must specify how these additional variables are to be held fixed when analytically continuing into the complex $s$-plane. 
For instance, one choice is to fix the scattering kinematics so that $\bfk_1 + \bfk_2 = 0$,  
enforcing the so-called \emph{centre-of-mass condition}, 
\begin{align}
\text{CM condition:} \qquad \omega_s = \sqrt{s} ~, ~ \omega_t=\omega_u = 0 \; ,
\label{eqn:CM}
\end{align}
since this corresponds to an incoming 2-particle state which is rotationally invariant, allowing the optical theorem \eqref{eqn:optical} to be simplified using the usual partial wave expansion. \eqref{eqn:CM} is the choice made in previous work \cite{Baumann:2011su, Baumann:2014cja, Baumann:2015nta}.
However, the resulting amplitude, $\mathcal{A}^{\rm (CM)}_s (s,t) = \mathcal{A}_s ( s,t, \sqrt{s},0,0)$ is not suitable for positivity arguments for two reasons: \\[5pt]
(i) $\mathcal{A}^{\rm (CM)}_s (s,t)$ is \emph{not} analytic in $s$ at fixed $t$ due to this non-analytic choice of $\omega_s$, which can be seen already in perturbation theory \footnote{
Explicitly, any interaction with an odd number of $n^\mu \partial_\mu$ derivatives may contain $\omega_s^{2n+1} =  s^n \sqrt{s}$ which has a square-root branch cut. 
Without an additional symmetry (e.g. time-reversal invariance) to enforce that $\omega_s$ appears only in even powers, in general the choice of kinematics \eqref{eqn:CM} does not produce an analytic $\mathcal{A} (s)$ when boosts are broken.
}, \\[5pt]
(ii) the condition \eqref{eqn:CM} is \emph{not} preserved under $s \leftrightarrow u$ crossing \eqref{eqn:crossing}, $\mathcal{A}_s^{\rm (CM)} (s,t) = \mathcal{A}_u (u,t,0,0,\sqrt{s}) \neq \mathcal{A}_u^{\rm (CM)} (u,t)$, and in fact \eqref{eqn:CM} is mapped to an unphysical kinematics (complex values of momenta). \\[5pt]
These two obstacles to constructing positivity bounds were pointed out as early as \cite{Baumann:2015nta}, and have thus far prevented the construction of robust bounds in EFTs without boost invariance. 
 
To overcome these issues, in this work we have identified a way to fix the energies which is analytic in $s$ and invariant under $s \leftrightarrow u$ crossing, so that the resulting amplitude is amenable to positivity arguments.  
We begin by noting that the conventional proofs of analyticity for Lorentz-invariant amplitudes (see e.g. the Appendix of \cite{deRham:2017zjm}) take place in the frame, $\bfk_1 - \bfk_3 = 0$, which corresponds to enforcing the \emph{Breit condition}, 
\begin{align}
&\text{Breit condition:}  \label{eqn:Breit}  \\
& \omega_s + \omega_u = \sqrt{4m^2-t} ~,~  \omega_s - \omega_u = \frac{s - u}{2 \sqrt{4m^2-t}} ~,~ \omega_t = 0 \; ,
 \nonumber
\end{align}
at small masses. This condition fixes the particle energies as analytic functions of $s$, and is preserved under $s \leftrightarrow u$ crossing (which swaps $s$ with $u$ and $\omega_s$ with $\omega_u$) so the resulting amplitude $\mathcal{A}_s^{\rm (Breit)} (s,t)$ enjoys a trivial crossing relation, $\mathcal{A}_s^{\rm (Breit)} (s,t) = \mathcal{A}_u^{\rm (Breit)} (u,t)$. 
For Lorentz-invariant theories, \eqref{eqn:CM} and \eqref{eqn:Breit} simply correspond to two different choices of Lorentz frame---one can therefore move freely from the centre-of-mass frame (in which unitarity is simplest) to the Breit frame (in which causality and crossing are simplest)---but when boosts are broken, $\mathcal{A}_s^{\rm (CM)} (s,t)$ and $\mathcal{A}_s^{\rm (Breit)} (s,t)$ are very different physical processes (different particle energies $\{ \omega_s, \omega_t, \omega_u\} $). 
For general kinematics, $\mathcal{A}_s (s,t,\omega_s, \omega_t, \omega_u)$, 
the energies do not typically satisfy the Breit condition \eqref{eqn:Breit}, 
however we can always parametrise them in the following way:
\begin{align}
\text{Breit parametrisation:} \hfill \label{eqn:BoostedBreit}  \\
 \omega_s + \omega_u &= 2 M \gamma   , ~~ \omega_s - \omega_u = \frac{s-u}{4M}   ~ , \qquad\qquad  \nonumber
\end{align}
where $\omega_s$ and $\omega_u$ have been written in terms of two new variables $\gamma$ and $M$ \footnote{
When $\omega_t = t = 0$, the physical $s$-channel region now corresponds to $s > 4m^2$, $\gamma \geq 1$, $M>0$.
}. Mathematically, \eqref{eqn:BoostedBreit} fixes the energies in a way which is analytic in $s$ (at fixed $t, M, \omega_t, \gamma$) and invariant under $s \leftrightarrow u$ crossing \footnote{
In fact, the energies could be fixed in a more general way, $\omega_s = f_s (s,t)$ and $\omega_u = f_u (s,t)$, and these would have the desired analyticity and crossing properties providing that $f_s$ and $f_u$ are analytic in $s$ at fixed $t$ and obey $ f_s (s,t)  \geq  \sqrt{ s +  f_u^2 (s,t) }$ and $f_u (u,t)  \geq \sqrt{ s + f_s^2 (u,t)  }$ for all real $s$ greater than $s_b$ (where the branch cut begins). However, we have found that it is the linear choice \eqref{eqn:BoostedBreit} which allows the closest connection between analyticity/causality and boundedness/locality (see Appendices~\ref{sec:analyticity_with_broken_boosts} and \ref{sec:sw}).  
}, and so the resulting amplitude $\tilde{\mathcal{A}}_s (s, t, M, \omega_t, \gamma)$ ($=\A_s$ with energies fixed as in \eqref{eqn:BoostedBreit}) enjoys all of the good properties of $\mathcal{A}_s^{\rm (Breit)} (s,t)$, in particular a trivial crossing relation, 
\begin{align}
\tilde{\mathcal{A}}_s (s, t, M, \omega_t, \gamma) =  \tilde{\mathcal{A}}_u (u, t, M , \omega_t, \gamma ) \; . 
\label{eqn:Breit_crossing}
\end{align}
Physically, varying $s$ with $\{ \gamma ,  M ,  \omega_t \}$ held fixed corresponds to changing the interaction energy while maintaining a constant velocity relative to the Breit frame (by contrast, the velocity of the centre-of-mass will vary with $s$). 
While it is clear that $\tilde{\A}_s$ is an analytic function of $s$ in the EFT, we further require that it is also analytic in the UV. In Appendix~\ref{sec:analyticity_with_broken_boosts}, we show that this requirement is related to the causality condition that operators commute outside of the $\pi$-cone.

~\\
{\bf (e) Polynomial Boundedness.}
Finally, we require a bound on the high-energy growth of the amplitude: 
\begin{align}
 \lim_{s\to\infty} |  \tilde{\A}_s ( s, t , M, \omega_t, \gamma ) |  <  \, s^2  \; , 
 \label{eqn:Froissart}
\end{align}  
where $s$ is taken large at fixed $\{ t, M, \omega_t, \gamma \}$.
In the Lorentz-invariant case, the classic results of Froissart and Martin \cite{PhysRev.129.1432,PhysRev.123.1053} show that polynomial boundedness of the amplitude,
$\lim_{s\to\infty} | \A_s (s,t) | < s^2$,
 follows from \emph{locality} (see e.g. \cite{Gribov:2003nw} for a modern textbook derivation).
In Appendix~\ref{sec:sw}, we argue that the analogous steps which lead to the Froissart-Martin bound in the Lorentz-invariant case can also be performed without Lorentz boosts (using the spherical wave expansion of \cite{Grall:2020tqc}), and so we will refer to property \eqref{eqn:Froissart} as locality of the UV physics.  



\section{Positivity Bounds}
\label{sec:positivity}

\noindent Armed with the above properties (a--e), we can now construct a dispersion relation for the amplitude $\tilde{\mathcal{A}}_s ( s, t, M, \omega_t, \gamma)$ and derive a number of new positivity relations which can be used to constrain low-energy EFTs in which boosts are broken. 

In the complex $s$-plane (with $\{t, M , \omega_t, \gamma \}$ held fixed), $\tilde{ \mathcal{A}}_s (s)$ has the same analytic structure as in a Lorentz-invariant theory---it is analytic everywhere except for the poles and branch cuts on the real $s$-axis that are required by unitarity. 
We can therefore follow the standard procedure \cite{Adams:2006sv}: evaluate the integral $\oint_C  d\mu  \,\tilde{ \mathcal{A}}_s (\mu ) / (\mu - s )^{n+1}$ first along a contour $C$ which closely encircles the pole at $\mu=s$ (giving $\partial_s^n \tilde{\mathcal{A}}_s$ by Cauchy's residue theorem), and then along a pair of semicircular contours in the upper/lower half-plane (closed above/below the real axis) whose radii go to infinity. These are equal by property (d) analyticity, which implies that,
\begin{align}
\frac{1}{n!} \partial_s^n \tilde{\mathcal{A}}_s (s)  &= C_{\infty} +  \int_{-\infty}^{\infty} \frac{d \mu}{\pi} \; \frac{ \text{Im} \, \tilde{\mathcal{A}}_s (  \mu  )   }{(\mu-s - i \epsilon)^{n+1} } 
\end{align}
where we have suppressed the dependence on $\{ t, M,\omega_t, \gamma \}$ and replaced $\tilde{\mathcal{A}}_s ( \mu + i \epsilon ) - \tilde{ \mathcal{A}}_s ( \mu - i \epsilon )$ with $2 i \text{Im} \, \tilde{\mathcal{A}}_s (\mu)$ using the real analyticity of $\tilde{ \mathcal{A}}_s$. 
$C_\infty$ is the contribution from the arcs at infinity, $C_{\infty} = \oint_{|\mu| \to \infty} \frac{d\mu}{\pi} \tilde{\mathcal{A}}_s (\mu) / \mu^{n+1}$, and vanishes by property (e) locality for all $n \geq 2$. 
Property (b) crossing can be used to relate the branch cut on the negative real axis to that along the positive real axis, so that there is a single integral $\int_{2m^2 - t/2}^{\infty} d \mu$ over the kernel,
\begin{align}
P_n ( \mu, s )  =  \frac{ \text{Im} \, \tilde{\mathcal{A}}_s (  \mu  )   }{(\mu-s)^{n+1} } -  \frac{ \text{Im} \, \tilde{\mathcal{A}}_u (  \mu )   }{(u - \mu )^{n+1} }   .
\label{eqn:Pn}
\end{align}
Following the philosophy of \cite{deRham:2017avq, deRham:2017imi} (see also \cite{Bellazzini:2017fep}), the strongest positivity bounds are obtained by subtracting as much of the IR (EFT) information as possible, so that the bounds represent only our ignorance of the UV. 
This corresponds to subtracting the portion of the branch cut up to the scale $s_b$ at which the EFT is no longer reliable,  
\begin{align}
\tilde{ \mathcal{A} }_s^{(n)} (s) := \frac{\partial_s^n \tilde{\mathcal{A}}_s (s) }{n!} - \int_{2m^2-t/2}^{s_b} \frac{d \mu}{\pi} P_n (\mu, s) \; .
\label{eqn:improv}
\end{align} 
$\tilde{ \mathcal{A} }_s^{(n)} $ is the $n^{\rm th}$ derivative of the EFT amplitude with low-energy branch cuts removed, and is related to the underlying UV completion (i.e. $\text{Im} \, \tilde{\A}_s (\mu)$ at large $\mu > s_b$) by the dispersion relation,
\begin{align}
\tilde{\A}_s^{(n)} (s)  =   \int_{s_b}^{\infty} \frac{d \mu}{\pi} P_n (\mu, s) \; .
\label{eqn:dispersion}
\end{align}

In the forward limit, $t \to 0$ and $\omega_t \to 0$ \footnote{
Note that since we are neglecting the mass, the different channels appear to collide when $t \to 0$ (since the $s$-channel, $s> 0$ and $\omega_s > \sqrt{s+\omega_u^2}$, becomes the complement of the $u$-channel, $s<0$ and $\omega_s < \sqrt{s+\omega_u^2}$). In principle these are always regulated by the small mass (which is necessary to establish the Froissart bound \eqref{eqn:Froissart}), and in practice once part of the branch cut up to $s_b$ has been subtracted there is always a gap through which to deform the integration contour. 
}, we show in Appendix~\ref{sec:sw} that $\text{Im} \, \tilde{A}_s$ coincides with $\text{Disc} \, \tilde{A}_s$, and therefore using the final property (c) unitarity we establish that $P_n (\mu, s) \geq 0$ for all $\mu \geq s_b$ and any even $n \geq 2$, providing that $| s | < s_b$ is within the EFT's regime of validity and that $M > 0$, $\gamma \geq 1$ have been chosen so that both cuts have physical kinematics.  
This gives our first positivity bound,
\begin{align}
\tilde{\mathcal{A}}_s^{(2N)} \big|_{\substack{t=0 \\ \omega_t = 0}} \geq 0  
\label{eqn:pos_fwd}
\end{align}
for all $N \geq 1$ and for any values of $s$, $M>0$ and $\gamma \geq 1$ which lie within the regime of validity of the EFT \footnote{
Note that the forward limit should be taken with $\omega_t \to 0$ \emph{first}, and then $t \to 0$. This ensures that there are no IR issues with the $t$-channel pole, which $\sim \omega_t^2/t$ in derivatively coupled theories.
}. This is the analogue of the positivity bounds $\partial_s^{2n} \mathcal{A} (s,t=0) > 0$ in Lorentz-invariant theories.

Note that we have \emph{not} required any notion of weak coupling to construct this dispersion relation---the EFT may be arbitrarily strongly coupled, as long it is able to capture the amplitude up to some scale $s_b$ which is larger than $2m^2-t$. 
At low energies, loop corrections to the EFT amplitude are suppressed by powers of $s/\Lambda^2$ and $\omega/\Lambda$, where $\Lambda$ is the EFT cutoff. 
In practice, the exact positivity bound \eqref{eqn:pos_fwd} therefore implies that,
\begin{align}
\frac{1}{2} \partial_s^{2} \tilde{\mathcal{A}}^{\rm tree}_s |_{  \substack{s=0 \\ t=0} }  \geq  \frac{2}{\pi} \int_{2m^2 -t}^{s_b} \frac{d \mu}{   \mu^3 } \; \text{Im} \, \tilde{\mathcal{A}}^{\rm 1-loop}_s |_{\substack{s=0 \\ t=0}}  
\label{eqn:pos_improv}
\end{align}
providing $s_b$ is chosen sufficiently low that higher order loop corrections can be ignored. 
We stress that subtracting up to $s_b$ with $M$ held fixed corresponds to energies $\omega_b/\Lambda \sim s_b/ M \Lambda$, and so the effective cutoff for the EFT amplitude $\tilde{\mathcal{A}}_s$ in the complex $s$ plane is set by the product $\Lambda M$. 
If the theory is weakly coupled, then we can trust the loop expansion even at energies near the cutoff (since there is an additional small parameter which suppresses these corrections), and so $\omega_b$ can be taken close to $\Lambda$, subtracting the entirety of the EFT branch cut (to any desired order in this weak coupling). 

For illustration we have focussed here on the forward limit $\omega_t =0$ and $t = 0$, which produces the infinite family of bounds \eqref{eqn:pos_fwd} that constrain every $\partial_s^{2N} \tilde{\mathcal{A}}_s$ with $N \geq 1$. In the Lorentz-invariant case, a strategy for placing analogous bounds on every $t$ derivative of $\mathcal{A}_s (s,t)$ was devised in \cite{deRham:2017avq}. We show in Appendix~\ref{app:more_pos_bounds} that, when boosts are broken, one can similarly go beyond the forward limit and place bounds on not just every even $s$ derivative of $\tilde{\mathcal{A}}_s$ beyond $\partial_s^2$, but also on \emph{every} $t$ derivative, $(\partial_t)^i$, and on \emph{every} energy derivative of the form $\left( \partial^2/\partial \omega_1 \partial \omega_3 \right)^j$. 
These bounds place highly non-trivial constraints on the low-energy EFT. 
In terms of the convenient variables $v = s-u$ and $\omega_- = \omega_s - \omega_u$, a general low-energy EFT amplitude can be expanded as,
\begin{align}
\mathcal{A}_s (s,t,\omega_s, \omega_t, \omega_u)  = \sum_{a, \,  b} C_{ab} \left( \omega_1, \, \omega_3 , \, \frac{ v }{\omega_- } \right)  \omega_-^{2a}  t^b
\end{align}
where we have made crossing symmetry \eqref{eqn:Breit_crossing} manifest. What we have achieved in equation \eqref{eqn:pos_fwd} is a bound on \emph{every} $C_{N0} ( \omega_1 , \omega_3, M) |_{\omega_1 = \omega_3}$ except $N=0$. Going beyond the forward limit in Appendix~\ref{app:more_pos_bounds} produces further bounds on \emph{every} symmetric derivative $C_{Ni}^{(j,j,0)} (\omega_1, \, \omega_3, \, M)|_{\omega_1 = \omega_3}$.


\section{Applications}
\label{sec:applications}

\noindent The positivity bound \eqref{eqn:pos_fwd} and its perturbative form \eqref{eqn:pos_improv} 
(as well as the bounds derived in Appendix~\ref{app:more_pos_bounds}) 
can be applied to \emph{any} low-energy EFT with rotation and translation invariance, and must be satisfied if the UV completion is to obey properties (c) unitarity, (d) causality and (e) locality (as defined above). To provide a concrete example of the power of these bounds, we will now focus on the simple EFT of a single scalar field $\pi$, 
to leading order in derivatives and up to quartic order in the field (we further assume an approximate shift symmetry so that the mass and potential interactions are small corrections). 
The action and corresponding $2 \to 2$ scattering amplitude are given by \cite{Grall:2020tqc},
\begin{widetext}
\begin{align} 
 S[ \pi ]&= \int d^4 x \, c_\pi^{-3} \Big(
-\tfrac{1}{2} (\partial\pi)^2
+ \frac{ \alpha_1 }{ \Lambda^2 }   \;  \dot \pi^3 
- \frac{ \alpha_2 }{ \Lambda^2 }  \; \dot \pi (\partial\pi)^2   
+ \frac{ \beta_1 }{ \Lambda^4 }  \; \dot \pi^4
- \frac{ \beta_2 }{  \Lambda^4 }   \; \dot \pi^2  (\partial\pi)^2  
+  \frac{ \beta_3 }{ \Lambda^4 }   \; (\partial\pi)^4  
\Big)\,,
\label{eqn:EFT_action}
\\[10pt]
\Lambda^4\A_s &= 
2 \beta_3(s^2+t^2+u^2)  
+ (2 \beta_2-4 \alpha_2^2) (s\,\omega_s^2 + t\, \omega_t^2 +u\, \omega_u^2) + 24 \omega_1 \omega_2 \omega_3 \omega_4  \left[   \left( \beta_1- 4 \alpha_1 \alpha_2  \right) 
-  \frac{3}{2} \alpha_1^2  \,   \left(\frac{\omega_s^2}{s}+\frac{\omega_t^2}{t}+\frac{\omega_u^2}{u}\right) \right] \; ,    \nonumber 
\end{align}
\end{widetext}
where $\dot \pi = n^\mu \partial_\mu \pi$ and $( \partial \pi )^2 = -\dot \pi^2 + c_s^2 \delta^{ij} \partial_i \pi \partial_j \pi$ is contracted using the effective metric which determines the free propagation \eqref{eqn:free_propagation}, $\{ \alpha_1, \alpha_2 ,\beta_1, \beta_2, \beta_3 \}$ are constant Wilson coefficients and the overall factor of the sound speed $c_\pi^3$ ensures canonical normalisation \cite{deRham:2017aoj, Grall:2020tqc}.
Despite its simplicity, this action captures the low-energy degree of freedom of a superfluid, and also the scalar fluctuations produced during inflation in the slow-roll decoupling limit.

~\\
{\bf Positivity Bounds.}
The only non-zero $\tilde{\A}_s^{(2N)}$ in the forward limit
at this order is $N=1$, which we write as $f = \Lambda^4 \tilde{\mathcal{A}}_s^{(2)} |_{\omega_t = 0, t=0}$. This is given explicitly by,  
\begin{align}
f (\gamma)  = 4 \beta_3 + 2 ( \beta_2 -2 \alpha_2^2 ) \gamma + \tfrac{3}{2} ( \beta_1 -4 \alpha_1 \alpha_2 ) \gamma^2 - \tfrac{9}{4} \alpha_1^2 \gamma^3 \; .    \label{eqn:amp_A2}  
\end{align}
%
The simplest positivity bound comes from \eqref{eqn:pos_fwd} with none of the EFT branch cut subtracted, which requires that $f \geq 0$ for all values of $\gamma \geq 1$ that lie within the regime of validity of the EFT.  
Setting $\gamma = 1$ happens to coincide with the centre-of-mass frame bound of \cite{Baumann:2015nta}, however we have derived the bound \eqref{eqn:amp_A2} in a different way which guarantees the required analyticity and positivity properties of the amplitude. 
In \cite{Grall:2020tqc}, it was shown using perturbative unitarity that $\gamma$ cannot be made arbitrarily large when $\alpha_1 \neq 0$, and here \eqref{eqn:amp_A2} provides an alternative demonstration of this from perturbative analyticity \footnote{ 
Physically, the presence of a $\dot \pi^3$ interaction allows for a loop-level exchange process which dominates the tree-level amplitude at large centre-of-mass velocities---this may also be related to the recent observation that such an interaction cannot be consistently coupled to gravity on a Minkowski background \cite{Pajer:2020wnj}.
}.

The one-loop contribution to the positivity bound \eqref{eqn:pos_improv} can be evaluated from the tree-level amplitude in \eqref{eqn:EFT_action} using the optical theorem, and is a simple polynomial in $s/M^2$ when $\gamma = 1$ and $\omega_t^2  = t  = 0$. At large $s$, 
\begin{align}
 \text{Im} \, \tilde{\mathcal{A}}^{\rm 1-loop}_s  = \frac{\beta^2}{ 4096 \pi} \; \frac{s^6}{     \Lambda^8 M^4}  \left( 1 + \mathcal{O} \left( \frac{M^2}{ s  } \right) \right) \; , 
 \label{eqn:A1loop_optical}
\end{align}
where $\beta^2 = \tfrac{2}{5} \left(  \beta_1 + \beta_2  - \left(  \frac{3}{2} \alpha_1 + \alpha_2 \right) \left(  \frac{\alpha_1}{2} + \alpha_2 \right)   \right)^2$.
Subtracting this branch cut up to an energy scale $\omega_b = s_b /4M \gg M$ gives a loop-improved positivity bound,
\begin{align}
 f |_{\gamma = 1}  \geq \frac{\beta^2}{16 \pi^2} \, \frac{ \omega_b^4 }{\Lambda^4} \left( 1 +  \mathcal{O} \left( \frac{M}{\omega_b}  \right) \right)  \; , 
 \label{eqn:pos_3}
\end{align}
which is strictly stronger than the previous bound. 
Note that if the EFT is strongly coupled, one cannot take $\omega_b$ close to $\Lambda$ without perturbation theory breaking down, and so \eqref{eqn:pos_3} is only ever a small improvement. 
However for weakly coupled theories there is an additional small coupling which suppresses higher order corrections even when $\omega_b$ is taken close to $\Lambda$, and in that case \eqref{eqn:pos_3} can be a significantly stronger bound, as we will now see explicitly in the case of a superfluid.

~\\
{\bf Superfluid.}
When describing the low-energy fluctuations of a superfluid, in which boosts are broken \emph{spontaneously} by the preferred spacetime direction $n^\mu$, the scalar $\pi$ corresponds to a Goldstone boson which transforms non-linearly under the broken boost symmetry \footnote{
In this context $\pi$ also transforms non-linearly under time translations, in which case any small breaking of its shift symmetry also corresponds to a small breaking of time translation invariance (see e.g. \cite{Finelli:2018upr}). 
},
\begin{align}
 n^\mu &\to n^\mu + b^\mu  ~ ,  &\pi &\to \pi + \Lambda^2 \;  b^\mu x^\nu \eta_{\mu\nu}
 \label{eqn:pi_boosts}
\end{align}
where $b^\mu n^\nu \eta_{\mu\nu} = 0$ is an infinitesimal space-like shift and $\Lambda$ is the energy scale associated with the symmetry-breaking \footnote{
In this context $\Lambda$ is often referred to as $f_\pi$ by analogy with the pion decay constant.
}.  
Note that while the low-energy action \eqref{eqn:EFT_action} depends only on the effective metric $Z^{\mu\nu}$ which determines the $\pi$ propagation, the UV boost symmetry introduces a new metric $\eta_{\mu\nu}$. They are related by $Z^{\mu\nu} = c_s^2 \eta^{\mu\nu} - (1 - c_s^2 ) n^\mu n^\nu$, where $c_s = c_\pi /c$ is the dimensionless ratio of the low-energy sound speed and the invariant UV speed preserved by \eqref{eqn:pi_boosts}. 
This non-linear symmetry fixes all but one Wilson coefficient at each order, and the coefficients appearing in \eqref{eqn:EFT_action} can all be fixed in terms of $\{ c_s, \alpha_1, \beta_1 \}$ only, 
\begin{align}
	\alpha_2= \frac{1-c_s^2}{2c_s^2}\,,\quad \beta_2=\frac{3}{2c_s^2}\alpha_1+\frac{(1-c_s^2)^2}{2c_s^4}\,,\quad \beta_3=\frac{1-c_s^2}{8c_s^4}\, .
	\label{eqn:superfluid}
\end{align} 

The simplest positivity bound $f \geq 0$ can be written in the form $\beta_1 \geq \beta_1^{\rm min} ( c_s, \alpha_1 )$, where,
\begin{align}
 \beta_1^{\rm min} = 
\frac{ 3 \alpha_1^2 \gamma }{2} 
+ \frac{ 2 \alpha_1 }{c_s^2} \left( 1 -c_s^2 - \frac{1}{\gamma} \right) 
- \frac{ 4 ( 1 - c_s^2) }{ 12  c_s^4 \gamma^2 } \; . 
\label{eqn:pos_1_superfluid}
\end{align}
Physically, this means that if the low-energy system has a sound speed $c_s$ and cubic interaction $\alpha_1$, then the underlying UV physics \emph{must} also produce a sufficiently large quartic interaction $\beta_1$ if it is to be consistent (i.e. unitary, causal, local). It is not possible, no matter how ingenious our UV model-building, to arrange for large effects at quadratic and cubic order in $\pi$ without also generating large effects at quartic order (unless we sacrifice one of the above fundamental properties).  

Including the one-loop correction to the positivity bound gave \eqref{eqn:pos_3}, which crucially depends non-linearly on $\beta_1$. 
In particular, there are some values of $\alpha_1$ and $c_s$ for which there are \emph{no} allowed values for $\beta_1$--- 
the improved bound \eqref{eqn:pos_3} can only be satisfied if,
\begin{align}
\left( 1 - c_s^2  + \frac{3}{2} \alpha_1 c_s^2 \right) \frac{ \omega_b^4 }{\Lambda^4} + 30 \pi^2 c_s^4  \geq 0 \; , 
 \label{eqn:pos_3_superfluid}
\end{align}
which allows us to bound $\alpha_1$ in terms of $c_s$ only---for instance for sufficiently small $c_s$ and large $\alpha_1 \sim 1/c_s^2$, this bound requires $\alpha_1 c_s^2 \gtrsim 2/3$. 

Finally, for the DBI tuning $\alpha_1 = \beta_1 = 0$, the positivity bound \eqref{eqn:pos_3_superfluid} can only be satisfied if $c_s \leq 1$, in which case it requires that new physics must become important at or below the scale $\omega_b^4 =  320 \pi^2 \Lambda^4 c_s^4 / ( 1 - c_s^2)^3$. 

\begin{figure*}[t]
\includegraphics[width=0.45\textwidth]{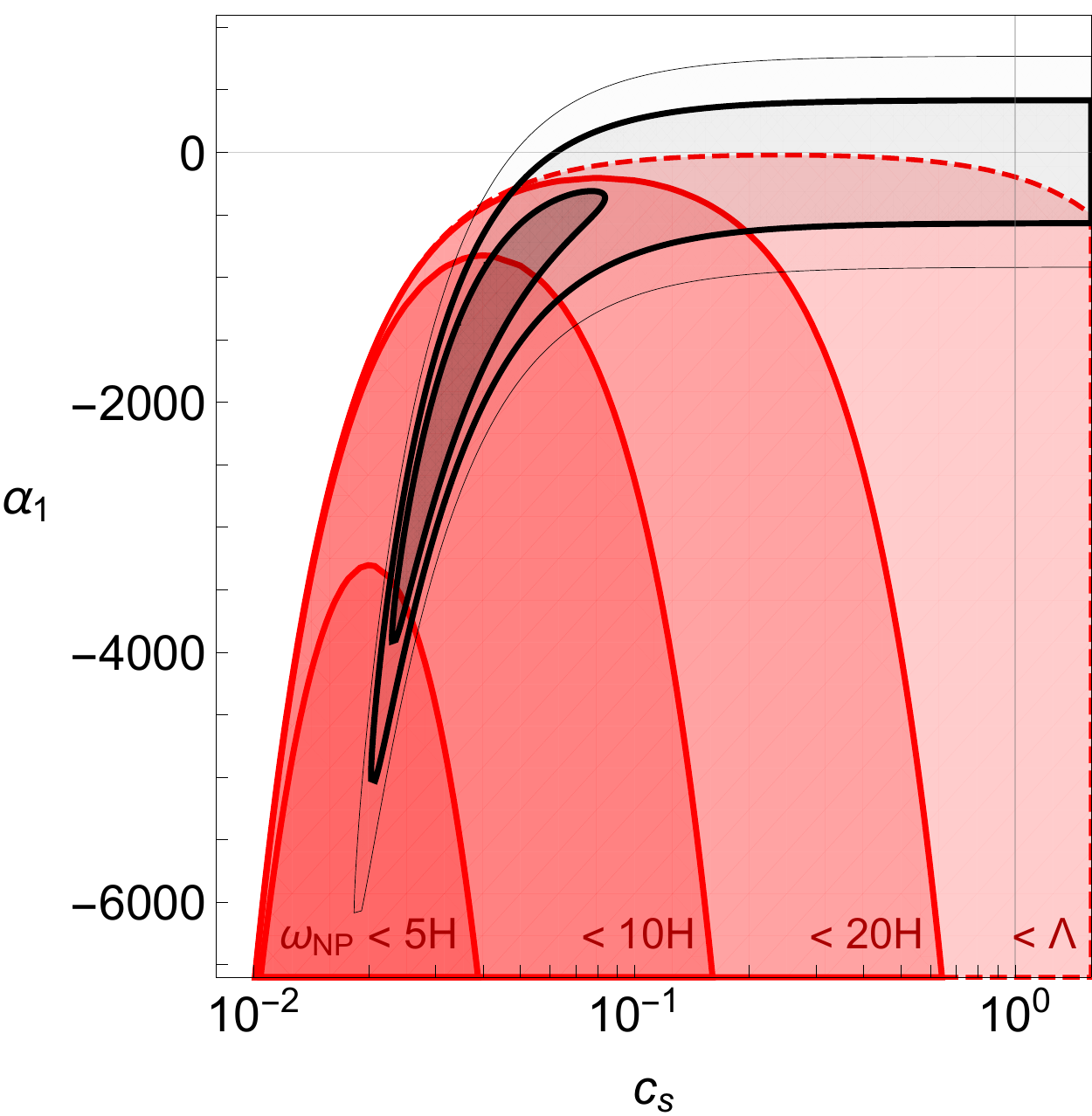} \qquad
\includegraphics[width=0.495\textwidth]{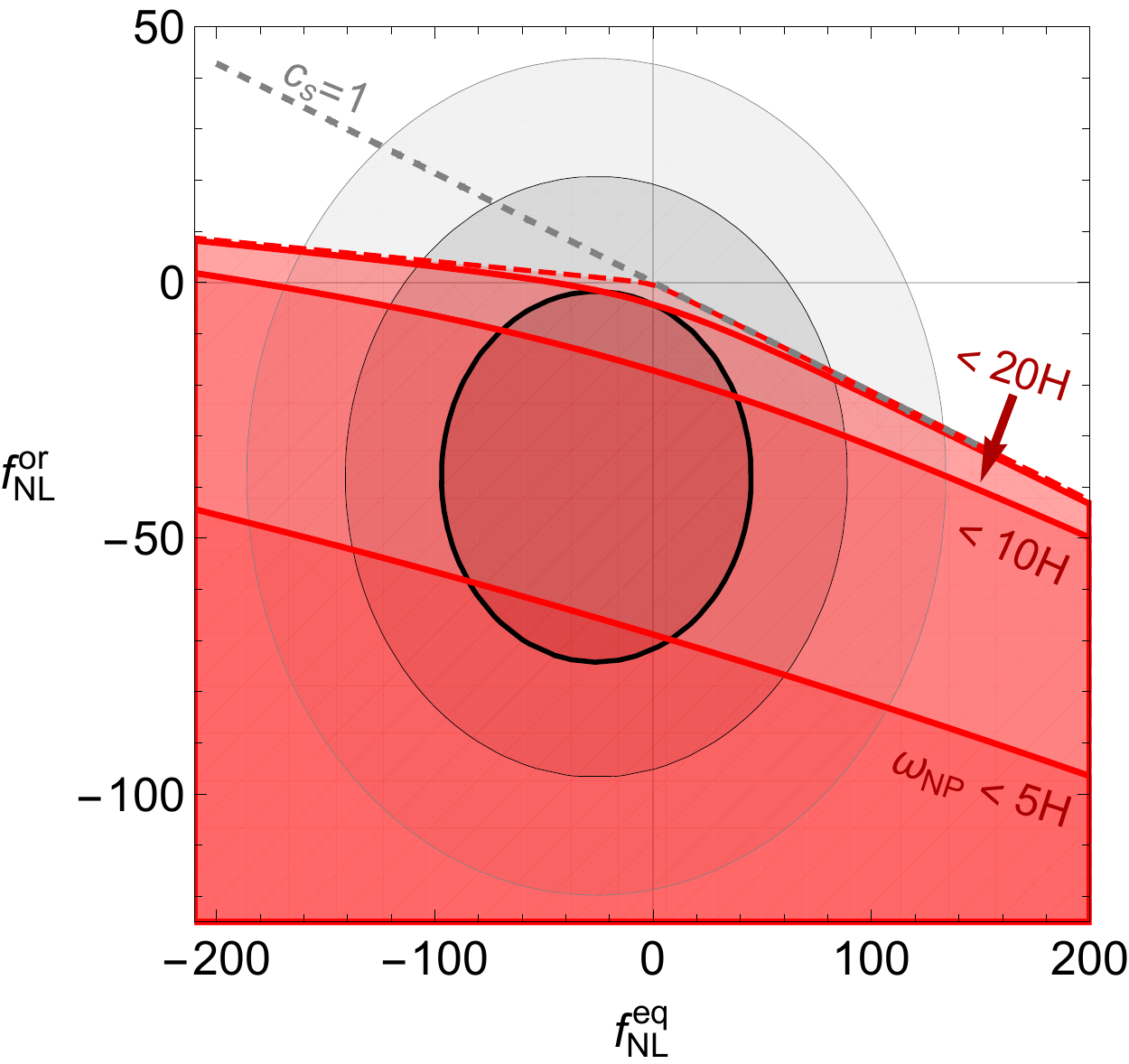}
\caption{
	Assuming that the UV completion is unitary, causal and local, our positivity bound \eqref{eqn:pos_3_superfluid} sets a scale $\omega_{\rm NP}$ beyond which inflation can no longer be single-field and weakly coupled. The red shaded regions correspond to $\omega_{\rm NP} < 5H, \, 10H, \, 20H$ and $\Lambda = 58.89 H$, and the grey contours show the $68\%, 95\%$ and $99.7\%$ confidence intervals from the Planck 2018 observations of the equilateral and orthogonal bispectrum $(f_{\rm NL}^{\rm eq}$ and $f_{\rm NL}^{\rm or}$), which are determined by the Wilson coefficients ($c_s$ and $\alpha_1$) appearing in the EFT of Inflation \eqref{eqn:EFT_action}. 
	For the majority of the $68\%$ confidence interval, the leading-order EFT requires the inclusion of new UV physics beyond the scalar fluctuations $\pi$ (or a non-perturbative treatment of $\pi$ loops) at an energies below $10H$.   
		\label{fig:positivity}
}
\end{figure*}

~\\
{\bf Inflation.}
During inflation, the expanding spacetime background spontaneously breaks temporal diffeomorphisms. Following the seminal work of \cite{Cheung:2007st}, these time diffeomorphisms can be restored (non-linearly realised) by introducing a single scalar degree of freedom, $\pi$. The resulting low-energy action for scalar and metric fluctuations is invariant under a gauged version of \eqref{eqn:pi_boosts} and in the decoupling and subhorizon limits,
\begin{align}
\text{Decoupling:}\; ~ \omega \gg \frac{\Lambda^2}{M_P} \; ,   \;\; \text{Subhorizon:} \;~ \omega \gg  H \; ,
\label{eqn:subhorizon}
\end{align}
where $H$ is the background Hubble rate, the scalar sector of the theory is described by \eqref{eqn:EFT_action} with the tuning \eqref{eqn:superfluid}.  
In this context, $c_s$ is the sound speed of scalar perturbations produced during inflation, and $\alpha_1$ ($\beta_1$) controls the primordial bi- (tri-)spectrum.   
The Planck \cite{Akrami:2018vks} measurements of the primordial power spectrum fix the symmetry-breaking scale to be $\Lambda = (58.89 \pm 0.21 ) H$ (at $68\%$ confidence), and the observed limits on the primordial bispectrum (i.e. on $\{ c_s, \alpha_1 \}$) are shown in Figure~\ref{fig:positivity}. 

The positivity bounds derived here provide a qualitatively new way to analyse these primordial signals. For instance, equation~\eqref{eqn:pos_1_superfluid} for $\beta_1^{\rm min}$ is a robust lower bound on the size of the trispectrum $g_{\rm NL}$, given measurements of the bispectrum $f_{\rm NL}$ and $c_s$. 
Taking for illustration the central values of the Planck 2018 contours \cite{Akrami:2018vks}, $f_{\rm NL}^{\rm eq} \approx -26$ and $f_{\rm NL}^{\rm or} \approx - 38$ ($c_s \approx 0.031$ and $\alpha_1 \approx -2125 $), the bound $f \geq 0$ gives $g_{\rm NL} \geq 12 \, 400$ ($\beta_1 \geq +6.4 \times 10^6$) \footnote{
This corresponds to setting $\gamma = 1$ in \eqref{eqn:pos_1_superfluid}.
This bound can be improved by using a larger value for $\gamma$, but one must ensure that it corresponds to kinematics which are both subhorizon \eqref{eqn:subhorizon} and within the resolving power of the EFT---see e.g. the cutoffs identified in \cite{Grall:2020tqc}. 
}. Comparing this with the currently allowed observational range, $-2.7 \times 10^6 < g_{\rm NL} < +1.1 \times 10^6$, we see that positivity arguments can be used to rule out over $70\%$ of the observationally allowed parameter space! Of course, since these numbers are sensitive to the relatively large uncertainties in $c_s$ and $\alpha_1$, they should be taken only as an illustration: the point is that, in future, as our measurements of the bispectrum improve, positivity bounds will provide a new way to ``bootstrap" information about the trispectrum (without the need to directly detect $g_{\rm NL}$).

Finally, our bound \eqref{eqn:pos_3_superfluid} can be used to place a restriction on the values of $\alpha_1$ and $c_s$ which are compatible with the UV properties listed in Section~\ref{sec:properties} (namely unitarity, causality and locality). 
In order to compare this bound with Planck observations, we must specify a value for $\omega_b$, the energy scale up to which we can reliably subtract the $\text{Im} \, \tilde{\mathcal{A}}_s$ from the dispersion relation. 
Rather than simply fix a value for $\omega_b$, instead we use \eqref{eqn:pos_3_superfluid} to define a scale $\omega_{\rm NP} (c_s, \alpha_1 )$, which is the largest possible energy at which \eqref{eqn:pos_3_superfluid} can be satisfied. 
This is the scale at which either New Physics (new degrees of freedom beyond $\pi$) or Non-Perturbative effects (e.g. a resummation of $\pi$ loops) \emph{must} become important, if we are to preserve unitarity, causality and locality.
We show in Figure~\ref{fig:positivity} how this scale compares with current Planck constraints on $\{ c_s, \alpha_1 \}$. The majority of the $68\%$ confidence region requires new physics within an order of magnitude of $H$.




~\\
{\bf Closing Remarks.}
In summary, we have constructed the first set of positivity bounds---constraints placed on a low-energy Effective Field Theory by the consistency of its underlying UV completion---when boosts are broken. 
The key to overcoming previous obstructions, namely a lack of analyticity and crossing symmetry, lay in identifying a new parametrisation for the amplitude---in our Breit variables, the $2 \to 2$ scattering amplitude $\tilde{\mathcal{A}}_s (s)$ enjoys the same structure as a Lorentz-invariant amplitude, and hence its dispersion relation can be used to bridge between the EFT and the UV completion. 
The new positivity bounds,  which constrain all $s$, $t$ and $\omega_1 \omega_3$ derivatives of the EFT amplitude, can be applied to a much wider class of systems than their Lorentz-invariant predecessors, and in particular are applicable to condensed matter and cosmology. 
As an example, we have shown how these bounds can be used to infer the scale of new physics which must be present in the early Universe, given the current Planck constraints on primordial non-Gaussianity. 
In addition to placing the positivity arguments on a more rigorous footing and deriving a number of new bounds, by overcoming the lack of crossing symmetry in \cite{Baumann:2015nta} we have been able to subtract the 1-loop EFT branch cut for the first time using the optical theorem (this is not possible using centre-of-mass kinematics since then the $u$-channel cut is not physical). 
The resulting bound \eqref{eqn:pos_3} is not only quantitatively stronger, but it gives a qualitatively new constraint on $\alpha_1$ in terms of $c_s$ alone \eqref{eqn:pos_3_superfluid}, providing a robust upper bound on the scale of new physics which is insensitive to the trispectrum. 
 
Although we focussed on examples in which Lorentz boosts were broken spontaneously, a UV Lorentz symmetry was not required to translate the properties (a-e) into bounds on the EFT.
Our positivity bounds should still apply in cases where the UV completion does not recover Lorentz invariance providing that property (d) still holds (this could be the case when there is still some maximum speed limiting the transfer of information, but which may not be the same for all observers). 
At the other extreme, in the event that the new physics which UV completes the EFT at its $\Lambda$ completely restores Lorentz invariance, then at weak coupling one could simply subtract all of the Lorentz-violating effects in the EFT and assume that the remaining $\tilde{A}_s^{(2n)}$ exhibits Lorentz-invariance: this could salvage kinematics such as the centre-of-mass frame, and potentially produce further positivity bounds. 
We have not pursued this strategy here because assuming symmetry restoration exactly at the EFT cutoff is a rather strong assumption about the UV completion (certainly much more restrictive than the assumptions of unitarity, causality and locality) \footnote{
For instance, the EFT could be UV completed before Lorentz invariance is restored, e.g. with a change in the propagation \eqref{eqn:free_propagation} \cite{Baumann:2011su} or with additional non-Lorentz invariant degrees of freedom \cite{Endlich:2013vfa}.
}.
 

The results presented here open up several avenues for future work. 
First and foremost, there are a number of systems in which Lorentz boosts are broken and yet the free propagation at low energies has the form $\omega = c_\pi k$, and our bounds can be immediately applied to the corresponding low-energy EFTs (see e.g. \cite{Nicolis:2011pv,Nicolis:2013sga,Endlich:2013vfa,Nicolis:2015sra,Delacretaz:2015edn,Pajer:2018egx, Alberte:2020eil,Grall:2020ibl})---in particular for the EFT of inflation, where we have explored but the simplest positivity bound, in future a systematic application of all of the available bounds should be carried out. 
Second, in covariant theories which possess both a Lorentz-invariant and a Lorentz-breaking background (e.g. $\langle \phi \rangle = 0$ and $\langle \phi \rangle \propto t$), it would be interesting to compare the bounds one would infer from the standard Lorentz-invariant arguments about the Lorentz-invariant background from the bounds derived here about the Lorentz-breaking background \footnote{
A good case study would be galileid theories \cite{Nicolis:2015sra}---the broken phase of Galileon EFTs---since around the Minkowksi background the Galileon (and Weakly Broken Galileon) is known not to satisfy positivity bounds \cite{Tolley:2020gtv}.
}.
Third, on the conceptual side, while we have focussed on the scattering of scalar fields, the approach put forward here (first identifying the correct frame in which analyticity and crossing are manifest) can also be extended to the scattering of fields of arbitrary spin. 
Fourth, in the context of inflation, in order to go beyond the subhorizon limit and fully account for the expanding FLRW background, one needs to move away from scattering amplitudes (which are no longer well-defined)---a natural candidate with which to connect UV and IR are the \emph{wavefunction coefficients}, and recent work has begun to shed light on how these objects are constrained by unitarity and locality \cite{Arkani-Hamed:2018kmz, Baumann:2019oyu, Baumann:2020ksv, Baumann:2020dch, Cespedes:2020xqq, Goodhew:2020hob, Pajer:2020wxk}.

\begin{acknowledgments}
\noindent We would like to thank Simon Caron-Huot, Claudia de Rham, Daniel Green, Johannes Noller, Enrico Pajer, David Stefanyszyn and Andrew Tolley for useful discussions and comments, and also the organisers of the workshop \emph{Cosmology 2021: the Rise of Field Theory}. 
TG is supported by the Cambridge Trust, and SM by an UKRI Stephen Hawking Fellowship. 
This work was partially supported by STFC consolidated grant ST/P000681/1. 
\end{acknowledgments}

\appendix

\section{Microcausality and Analyticity} 
\label{sec:analyticity_with_broken_boosts}

\noindent 
Analyticity of the scattering amplitude is fundamentally tied to the principle of causality. 
For Lorentz-invariant theories this connection is well-understood and analyticity of the Lorentz-invariant scattering amplitude $\A_s (s,t)$ at fixed $t$ is well-established \cite{Bremermann:1958zz, bogoliubov1959introduction, Martin:1962rt, Hepp_1964, Bros:1964iho, Jin:1964zza, Martin:1965jj, Mahoux:1969um}. 
 
When boosts are spontaneously broken, far less is known about the analytic structure of $\mathcal{A}_s (s,t,\omega_s, \omega_t, \omega_u)$.
In this appendix, we will show how the requirement of microcausality is closely related to the analyticity property (d) described in Section~\ref{sec:properties}. In particular, we will show that whenever the UV interactions preserve/contract the causal cone (i.e. $c_{\rm UV} \leq c_\pi$), then $\mathcal{A}_s (s,t, \omega_s, \omega_t, \omega_u)$ is guaranteed to be analytic for certain domains in $\{ s, \omega_s - \omega_u \}$ at fixed $\{ t, \omega_s + \omega_u , \omega_t \}$.  
When UV interactions expand the caucal cone (i.e. $c_{\rm UV} > c_\pi$), then the conventional microcausality arguments are not strong enough to guarantee analyticity, but we can nonetheless show that $\A_s (s,t,\omega_s, \omega_t,\omega_u)$ is analytic in the $\{\omega_s, \omega_t, \omega_u \}$ at fixed $\{ s, t \}$ at any order in perturbation theory. 
 
~\\
{\bf Microcausality.}
In a quantum theory, causality (signals cannot propagate faster than some speed $c_{\rm max}$) is encoded in the operator relation,
\begin{align}
 [  \mathcal{O} (x) ,  \mathcal{O} (y) ] = 0 \;\; \text{when} \;\; c_{\rm max} \, | x^0 - y^0 | <  | \bfx - \bfy | \; , 
 \label{eqn:causality}
\end{align}
which is the familiar statement that operators (observables) at space-like separated points commute. In order to connect this requirement to the amplitude, we define the position-space current operator,
\begin{equation}
	J_A (x) \equiv i  \frac{\delta \hat S}{\delta \pi_A (x)}{\hat{S}}^{\dagger} \; , 
\end{equation}
so that the amplitude can be written as,
\begin{align}
	\mathcal{A}_s  &=\int\d^4 x \;  e^{-\frac{i}{2} k_{24 \, \mu} x^\mu } \Theta(x^0) 
	\matrixel{ \bfk_3}{\comm{ J_{\bar 4} \left( \frac{x}{2}  \right) }{J_2 \left( - \frac{x}{2} \right) }}{ \bfk_1}  ,
	\label{eq:matrixel_schannel}
\end{align}
where $k_{24 \, \mu }  = k_{2 \, \mu }  + k_{4\, \mu}$ (where now we use conventions in which $\omega_4 > 0$, in contrast to the main text). 
When the momenta $\bfk_1$ and $\bfk_3$ labelling the one-particle states are both real, then causality \eqref{eqn:causality} requires that this matrix element vanishes unless $x$ is time-like, i.e. the integration region can be restricted to $c_{\rm max} \, x^0 > | \bfx |$. 
If $k_{24 \, \mu}$ is analytically continued into the complex plane (with $\bfk_1$ and $\bfk_3$ fixed), this integral representation for the amplitude converges whenever $\text{Im} \left[ k_{24 \, \mu} x^\mu  \right]  < 0$ for every real $x^\mu$ in this range. The strongest requirement corresponds to $\hat{\bfx}$ aligning with $\text{Im} \, \bfk_{24}$, and the amplitude is therefore analytic for any complexification of $k_{24 \, \mu}$ which obeys,
\begin{align}
 \text{Im} \, \omega_{24}  \geq c_{\rm max} \, | \text{Im} \, \bfk_{24} | > 0 \; ,
\label{eqn:k24cond}
\end{align}
as a consequence of \eqref{eqn:causality}. 
This is the connection between causality (i.e. space-like separated operators must commute) and the analyticity of the amplitude for complex values of momenta which we will exploit---when written in terms of the variables $\{ s,t,u, \omega_s, \omega_t, \omega_u \}$, \eqref{eqn:k24cond} describes a domain in which $\mathcal{A}_s$ must be analytic for causal interactions.

~\\
{\bf With Boosts.}
Before discussing theories without boost invariance, let us briefly review how \eqref{eqn:k24cond} is used to establish analyticity in $s$ at fixed $t$ in the Lorentz-invariant case.
For a Lorentz-invariant scattering amplitude, one can use boosts to go to the Breit frame, in which $\bfk_{13} = 0$ (i.e. $\bfk_1 = -\bfk_3$). Conservation of momentum then requires that $\bfk_1 \cdot \bfk_{24} = 0$ and so $\bfk_{24}$ is orthogonal to both $\bfk_1$ and $\bfk_3$. For instance, using rotations to align $\bfk_t = \bfk_1 - \bfk_3$ along the $z$-axis, the particle momenta can be parametrised in this frame by,
\begin{align}
 \bfk_1 = \left( \begin{array}{c} 
 0 \\
 0 \\
 \frac{k_t}{2}
 \end{array} \right) ~ , \;\;   
 \bfk_2 = \left( \begin{array}{c} 
 \tfrac{k_{24}}{2}  \\
 0 \\
 - \frac{k_t}{2}
 \end{array} \right) ~ , \;\;
 \bfk_3 = \left( \begin{array}{c} 
 0 \\
 0 \\
 -\frac{k_t}{2}
 \end{array} \right)  ~ , \;\; 
\end{align}
where $\bfk_4 = \bfk_1 + \bfk_2 - \bfk_3$ is fixed by momentum conservation. 
 The two independent variables, $k_t = | \bfk_1 - \bfk_3 |$ and $k_{24} = | \bfk_2 + \bfk_4|$, are related to the Mandelstam variables by $c^2 k_t^2 = -t$ and $\omega_{24} \sqrt{4m_1^2 - t} = s + \frac{t}{2} - m_1^2 - m_2^2$, where $\omega_{24} = \sqrt{ c^2 k_{24}^2 - 4 m_2^2 -t } $ is the energy appearing in \eqref{eqn:k24cond}. 
These are often known as \emph{Breit coordinates}, and have the advantage that analytically continuing $k_{24}$ into the complex plane with $k_t$ held fixed (i.e. the complex $s$ plane at fixed $t$) deforms particles 2 and 4 into the complex plane while keeping $\bfk_1$ and $\bfk_3$ real.  
Analysing the precise domain of analyticity is an involved problem, and in particular requires careful treatment of the square roots in $k_{24} (s,t)$. 
However, when considering $t$, $m_1$ and $m_2$ to be much smaller than $|s|$, we have that $\omega_{24} \approx  s/\sqrt{4m_1^2 - t} \approx c k_{24}$, and so the condition \eqref{eqn:k24cond} becomes simply $\text{Im} \, s > 0$ when $c_{\rm max} \leq c$ (i.e. when the maximum speed at which signals can propagate is contained within the lightcone which determines the propagation of free fields). 
Causality, for a Lorentz-invariant amplitude, therefore requires that $\mathcal{A}_s$ be an analytic function of $s$ in the upper half-plane when $t$ (and the masses) are fixed to real physical values.

~\\
{\bf Without Boosts.}
Without invariance under Lorentz boosts, it is no longer possible to set $\bfk_{13} = 0$ in general. Instead, using only invariance under rotations, we can express the momenta in terms of three magnitudes and three angles,
\begin{align}
 \bfk_{t} = k_t \; \hat{\mathbf{e}}_{ 0,0 } ~ , \;\;   
 \bfk_{13} =  k_{13} \; \hat{\mathbf{e}}_{ \theta_{13} ,0 } ~ , \;\; 
  \bfk_{24} = k_{24} \; \hat{\mathbf{e}}_{  \theta_{24} , \phi_{24} } ~ , 
\end{align}
where $\hat{\mathbf{e}}_{\theta, \phi} = \left( \sin \theta \cos \phi, \sin \theta \sin \phi, \cos \theta \right)$ is a spatial unit vector. 
To express $\{ k_t , k_{13} , k_{24} , \theta_{13}, \theta_{24}, \phi_{24} \}$ compactly in terms of $\{ s, t , u , \omega_s, \omega_t, \omega_u \}$, we will now assume that the masses are negligible. 
Then the variables $\omega_t$ and $t$ correspond to specifying $k_t$ and fixing the angles $\theta_{13}$ and $\theta_{24}$ in terms of the magnitudes $k_{13}$ and $k_{24}$, 
\begin{align}
c_\pi k_{t} = \sqrt{ \omega_t^2 - t} \, , \; \cos \theta_{13} = \sqrt{ \frac{ 1 - \frac{t}{c_\pi^2 k_{13}^2} }{ 1 - \frac{t}{\omega_t^2} } } \, ,  \; \cos \theta_{24} = \sqrt{ \frac{ 1 - \frac{t}{c_\pi^2 k_{24}^2} }{ 1 - \frac{t}{\omega_t^2} } } \; .
\end{align}
The variables $\omega_s$ and $\omega_u$ then correspond to $k_{13}$ and $k_{24}$, 
 \begin{align}
c_\pi  k_{13} &= \sqrt{ ( \omega_s + \omega_u )^2 + t  } \; , \;\;  &c_\pi k_{24} &= \sqrt{ ( \omega_s - \omega_u )^2 + t  } \; , 
 \end{align}
and finally $s$ determines the remaining parameter $\phi_{24}$, 
\begin{align}
\cos \phi_{24} = 
- \frac{ 2 s+t
   - \frac{t}{c_\pi^2 k_t^2} \left(  \omega_s^2  - \omega_u^2 \right) }{
c_\pi^2 k_{13} k_{24} \sin \theta_{13} \sin \theta_{24}    
   }
   \; . 
\end{align}
The advantage of this parametrisation is that $\bfk_1$ and $\bfk_3$ can be kept real by holding $\{ t, \omega_t ,  \omega_s + \omega_u \}$ fixed, while condition \eqref{eqn:k24cond} then places a restriction on the complex values which $\{ s,  \omega_s - \omega_u \}$ may take.
Just as in the Lorentz-invariant case, precisely characterising this region requires careful treatment of square roots of the form $\sqrt{ (\omega_s - \omega_u)^2 + t - \omega_t^2 }$. 
To illustrate the key features, we will consider both $t$ and $\omega_t^2$ to be fixed at values much smaller than $s$, $( \omega_s - \omega_u)^2$ and $(\omega_s + \omega_u)^2$. 
This simplifies \eqref{eqn:k24cond} to,
\begin{align}
c_\pi \text{Im} \left[ \omega_s - \omega_u \right] \geq 
c_{\rm max} \sqrt{ 
\text{Im} \left[  a  \right]^2 +
\text{Im} \left[ b  \right]^2     
}   > 0
\label{eqn:wsucond}
\end{align}
where,
\begin{align}
 a \approx \frac{\omega_t}{\sqrt{-t}} \frac{2s}{ \omega_s + \omega_u  }
\; , \;\; b \approx \sqrt{ (\omega_s - \omega_u)^2  - a^2 } \; . 
\end{align}
Once $t, \omega_t$ and $\omega_s + \omega_u$ are fixed to real physical values, then \eqref{eqn:wsucond} specifies the domain analyticity for $\omega_s - \omega_u$ and $s$. In particular, note that when $(\omega_s - \omega_u)^2 \gg s$ is large (or when $\omega_t \to 0$) this becomes simply $\text{Im} \left[ \omega_s - \omega_u \right] > 0$ when $c_{\rm max} \leq c_\pi$ \footnote{
When $c_\pi < c$, then we are allowing for UV interactions which \emph{expand} the $\pi$-cone ($x^0 = | \bfx| / c_\pi$) to the Minkowski lightcone ($x^0 = | \bfx | / c$). Analyticity is more difficult to prove non-perturbatively in that case since the usual arguments rely on the assumption that interactions do \emph{not} change the causal support of the free theory. 
}, and so it is $\omega_s - \omega_u$ which plays the role of $s$ from the Lorentz-invariant case. 
For our purposes, fixing $M$ and $\gamma$ to real values gives,
\begin{align}
 a \approx \frac{\omega_t}{\sqrt{-t}}  \; \frac{s}{ \gamma M } \; , \;\; b \approx \frac{s}{2M} \sqrt{ 1 + \frac{4 \omega_t^2}{ \gamma^2 t}  }
\end{align} 
and therefore \eqref{eqn:wsucond} becomes $\text{Im} \, s > 0$ when $c_{\rm max} \leq c_\pi$. When making $s$ complex at fixed (real) values of $\{ t ,  M , \omega_t , \gamma  \}$, causality \eqref{eqn:causality} requires that $\mathcal{A}_s$ is analytic in the upper half-plane, just as in the Lorentz-invariant case. Crucially however, the upper bound required of $c_{\rm max}$ is set by the the $\pi$-cone (i.e. the free propagation \eqref{eqn:free_propagation}). If one imagines a UV completion in which there is some maximum speed $c_{\rm UV}$, then causality in that UV completion is only strong enough to guarantee analyticity of the amplitude if $c_{\rm UV} \leq c_\pi$. 
In cases where $c_{\rm UV} > c_\pi$, a more sophisticated argument, which accounts for how the operators behave in the region between the two cones (i.e. $1/c_{\rm UV}  < x^0 / | \bfx | \leq 1/c_{\pi}$, where \eqref{eqn:causality} no longer guarantees commutation) is needed.

~\\
{\bf Crossing.}
To establish analyticity in the full complex $s$-plane, one needs to repeat the above argument in the $u$-channel. To see this, consider the $u$-channel amplitude corresponding to the process $\pi_1\bar\pi_{4}\to\pi_3\bar\pi_{2}$:
\begin{align}\nn
    \mathcal{A}_u  &=\int\d^4 x \;  e^{ \frac{i}{2} k_{24 \, \mu} x^\mu } \; \Theta(x^0) 
    \matrixel{ \bfk_3}{\comm{ J_{2} \left( \frac{x}{2}  \right) }{J_{\bar 4} \left( - \frac{x}{2} \right) }}{ \bfk_1}  ,\\
    &=-\int\d^4 x \;  e^{- \frac{i}{2} k_{24 \, \mu } x^\mu} \; \Theta(-x^0) 
    \matrixel{ \bfk_3}{\comm{ J_{\bar 4} \left( \frac{x}{2}  \right) }{J_{2} \left( - \frac{x}{2} \right) }}{\bfk_1} .
    \label{eq:matrixel_uchannel}
\end{align}
Analytically continuing $k_{24}^\mu$ into the complex plane, the above integral representation converges if $\Im[k_{24 \, \mu} x^\mu] < 0$ for all $-c_{\rm max} \, x^0 >  | \bfx |$, i.e. providing that
\begin{equation}
    -\Im  \omega_{24} \geq c_{\rm max} \, |\Im \bfk_{24}| > 0 \; . 
    \label{eqn:k24cond_uchannel}
\end{equation}
In the Lorentz invariant setting, for $t$, $m_1$ and $m_2$ much smaller than $|s|$, this implies $\Im s<0$ and so $\A_s$ is required to be analytic also in the lower half-plane $s$-plane. 
When boosts are broken the condition \eqref{eqn:k24cond_uchannel} reads
\begin{align}
- c_\pi \text{Im} \left[ \omega_s - \omega_u \right] \geq  c_{\rm max}
\sqrt{ 
\text{Im} \left[  a  \right]^2 +
\text{Im} \left[ b  \right]^2     
} > 0   \,,
\label{eqn:wsucond_uchannel}
\end{align}
which, in the limit of large $(\omega_s - \omega_u)$ (or $\omega_t\to0$) becomes $\text{Im} \left[ \omega_s - \omega_u \right]<0$ (when $c_{\rm max} \leq c_\pi$), so that $\mathcal{A}_s$ is analytic in the whole $(\omega_s - \omega_u)$ plane (except for possible branch cuts on the real axis $\text{Im} \left[ \omega_s - \omega_u \right] = 0$, which are fixed by unitarity in the physical regions).
In our boosted Breit coordinates, \eqref{eqn:wsucond_uchannel} describes analyticity in the lower-half of the complex $s$ plane ($\Im s<0$) with $\{s,t, M,\omega_t, \gamma\}$ fixed. $\tilde{\mathcal{A}}_s$ is therefore analytic in the whole complex $s$ plane, modulo possible poles/branch cuts on the real $s$ axis (which are fixed by unitarity), whenever $c_{\rm max} \leq c_\pi$.
Since this standard causality argument is not strong enough to prove analyticity non-perturbatively when $c_{\rm max} > c_\pi$, for that case we turn to perturbative arguments.

~\\
{\bf Perturbation Theory.}
Within perturbation theory, there can be \emph{no} poles or branch cuts in the energies $\{ \omega_s,\omega_t$, $\omega_u \}$ at fixed $s$ and $t$. 
To see this, note that we can always write $\omega_i=n^\mu p_{i,\mu}$ and then strip-off from the amplitudes all factors of $n^\mu$,
\begin{equation}
	\A_s^{L\text{-loop}} = n^{\mu_1}\dots n^{\mu_j} \A_{\mu_1\dots \mu_j }^{L\text{-loop}} \, .
	\label{eqn:A_strip}
\end{equation}
The remaining $\mathcal{A}^{\text{L-loop}}_{\mu_1 \dots \mu_j}$ is then a series of $L$ loop integrals over \emph{covariant} integrands (with respect to the $Z^{\mu\nu}$ which determines the free propagator), which can only ever produce non-analyticities in the invariants $s$ and $t$.

~\\
{\bf An Example.}
We illustrate this point by explicitly computing the 1-loop diagram generated by two $\dot\pi^4$ interactions. The $s$-channel diagram can be stripped as in \eqref{eqn:A_strip}, leaving behind,
\begin{align}\nn
	&\A_{\mu_1\dots \mu_4}^{\text{1-loop}} = \frac{4!^2 \beta_1^2}{2 \Lambda^8}\, \omega_1 \omega_2 \omega_3 \omega_4  
	\int \frac{\d^4 k}{(2\pi)^4}\frac{ q^{\mu_1} q^{\mu_2}k^{\mu_3}k^{\mu_4}}{\left(k^2-m^2 \right)\left( q^2-m^2 \right)}
\end{align}
where $q^\mu=p_1^\mu+p_2^\mu + k^\mu$. The $t$ and $u$-channel diagrams are the same up to a permutation of the external legs. 
In dimensional regularisation, this integral evaluates to, 
\begin{align}
	\A^{\text{1-loop}}_s &= \frac{3}{1280}\frac{\beta_1^2}{16\pi^2 \Lambda^8} F_s (\omega_j)  \log\left( -s \right)   	\label{eqn:A1loop_b1} \\
	&+ t,\,u\text{ permutations} + \text{local counter-terms}
	\nonumber 
\end{align}
where $F_s (\omega_j) =  \omega_1 \omega_2 \omega_3 \omega_4 \left(10\,\omega_s^4+\frac{15}{3}s\,\omega_s^2+\frac{15}{4}s^2\right) $ is an analytic function of the energies. 
As claimed above, the only branch cuts in this expression are in the invariant variables $s,t$ and $u$.
This feature, that the energies appear only as an analytic factor in each term, is generic. 

The explicit computation \eqref{eqn:A1loop_b1} also provides a useful cross-check of our earlier equation \eqref{eqn:A1loop_optical}.  
Using the optical theorem and the tree level amplitude \eqref{eqn:EFT_action}, the one-loop imaginary part from the $\beta_1$ interaction is found to be, 
 \begin{align}
 &\text{Im} \, \tilde{\mathcal{A}}^{\rm 1-loop}_s |_{ \substack{ \gamma = 1 \\ \omega_t = 0 \\ t= 0}} \nonumber \\
 &=  \frac{3}{640} \frac{\beta_1^2}{16 \pi} \; \frac{s^2}{  \Lambda^4 } 
  \frac{  256 M^8 + 384 M^6 s + 256 M^4 s^2 + 24 M^2 s^3  +  s^4}{ \tilde{\Lambda}^8 }\,,
\end{align}
where $\tilde{\Lambda}^2 = M \Lambda$. This agrees with the discontinuity across the logarithms in \eqref{eqn:A1loop_b1}.


\section{Spherical Wave Expansion} 
\label{sec:sw}

\noindent In this appendix we briefly review properties of the spherical wave expansion for amplitudes without Lorentz boosts introduced in \cite{Grall:2020tqc}, the positivity properties of $\text{Disc} \, \mathcal{A}_s$ and the Froissart-Martin bound \eqref{eqn:Froissart}. 

The key idea is to express the amplitude in a particular channel (e.g. the $s$ channel) not in terms of linear momentum eigenstates, $| \bfk_1 \rangle = | \omega_1 \vartheta_1 \varphi_1 \rangle$, but in terms of angular momentum eigenstates $| \omega_1 \ell_1 m_1 \rangle$ (where $m_1$ is the angular momentum conjugate to $\varphi_1$ and $\ell_1$ represents the total angular momentum, $\ell_1 (\ell_1+1)$). 
Mathematically, this amounts to the decomposition,  
\begin{align}
\mathcal{A}_{s} ( \bfk_1, \bfk_2, \bfk_3 ,\bfk_4 ) = \sum_{ \substack{ \ell_1 ,  \ell_3  \\ m_1, m_3 }} f_{\ell_1}^{m_1} ( \hat{\bfk}_1 ) f_{\ell_3}^{m_3 *} ( \hat{\bfk}_3 )  \, a^{m_1 m_3}_{\ell_1 \ell_3} ( \omega_s, s )
\label{eqn:sw_general}
\end{align}
where $f_{\ell }^{m}$ are a complete set of functions on the sphere. 
This makes rotational invariance manifest and in particular greatly simplifies the optical theorem \eqref{eqn:optical}.
Comparing with the corresponding spherical wave expansion for the $\bar{s}$-channel gives, 
\begin{align}
\text{Disc} \, \mathcal{A}_s = \sum_{ \substack{ \ell_1 ,  \ell_3  \\ m_1, m_3 }} f_{\ell_1}^{m_1} ( \hat{\bfk}_1 )  f_{\ell_3}^{m_3 *} ( \hat{\bfk}_3 ) \text{Disc} \, a^{m_1 m_3}_{\ell_1 \ell_3}  
\label{eqn:sw_Disc}
\end{align}
where $\text{Disc} \, a^{m_1 m_3}_{\ell_1 \ell_3}  = \tfrac{1}{2i} \left(  a^{m_1 m_3}_{\ell_1 \ell_3} -  a^{m_3 m_1 *}_{\ell_3 \ell_1}  \right)$. The optical theorem then takes the form,
\begin{align}
 \text{Disc} \, a^{m_1 m_3}_{\ell_1 \ell_3} =  \sum_n \; z^n_{\ell_1 m_1} z^{n *}_{\ell_3 m_3}
 \label{eqn:sw_unitarity}
\end{align}
where $z^n_{\ell_1 m_1} = \int d^2 \hat{\bfk}_1 \,  \bar{f}_{\ell_1}^{m_1} (\hat{\bfk}_1 ) \mathcal{A}_{\pi_1 \pi_2 \to n}$ (where $\bar{f}_{\ell}^{m}$ satisfies $\int d^2 \hat{\bfp} \bar{f}_{\ell}^{m} ( \hat{\bfp} ) f_{\ell'}^{m'} ( \hat{\bfp} ) = \delta_{\ell \ell'} \delta_{m m'}$) and the sum runs over all intermediate $n$ particle state (and includes integrals over their kinematics). Unitarity therefore requires that $ \text{Disc} \, a^{m_1 m_3}_{\ell_1 \ell_3}$ is a positive definite matrix in $\ell_1 m_1$ and $\ell_3 m_3$.

~\\
{\bf Forward Limit.}
Taking the forward limit ($\hat{\bfk}_1 = \hat{\bfk}_3$) in  \eqref{eqn:sw_Disc} therefore immediately establishes $\text{Disc} \, \mathcal{A}_s \geq 0$. 
Furthermore, in this limit we have that the $f_{\ell}^m$ can be replaced with their moduli,
\begin{align}
 \mathcal{A}_s |_{\hat{\bfk}_1 = \hat{\bfk}_2} = \sum_{ \substack{ \ell_1 ,  \ell_3  \\ m_1, m_3 }} | f_{\ell_1}^{m_1} ( \hat{\bfk}_1 ) | \; | f_{\ell_3}^{m_3} ( \hat{\bfk}_1 ) | \;  a^{m_1 m_3}_{\ell_1 \ell_3} 
\end{align}
which proves that $\text{Disc} \, \A_s$ and $\text{Im} \, \A_s$ coincide in the forward limit. 
To prove the positivity of $\text{Disc} \, \A_s$ beyond the forward limit, one needs to specify how the angles in $\hat{\bfk}_1$ and $\hat{\bfk}_3$ are related to the analytic variables $\{ s, t, \omega_s, \omega_t, \omega_u \}$.

~\\
{\bf Angular Variables.}
Once the total incoming energy $\omega_s$ and momentum $k_s = | \bfk_1 + \bfk_2 |$ are fixed, the allowed frequencies for $\omega_1$ lie on an ellipse with semimajor axis $\omega_s/2$ and eccentricity $\rho_s = c_\pi k_s/\omega_s$. 
These can be parametrised with either the angle $\theta_1$ between $\bfk_1$ and $\bfk_s$ (the true anomaly of the  ellipse), or alternatively by the angle $\vartheta_1$ (the eccentric anomaly of the ellipse), which for small masses are related by \footnote{
At energies which are comparable to the mass of $\pi$, \eqref{eqn:psi_def} becomes,
\begin{align}
 \frac{\omega_j - \rho_s k_j}{1 - \rho_s} = \frac{\omega_s}{2} \left( 1 - \rho_s \cos \psi_j \right) \; , 
\end{align}
where $k_j = \sqrt{\omega_j^2-m_j^2}$. This can be inverted for $\cos \psi_j$ straightforwardly, although now $\cos \psi_1 + \cos \psi_2$ no longer vanishes (it is instead proportional to the mass) and so the Bose symmetry condition \eqref{eqn:sw_symm} is more involved---it related $a^{m_1 m_3}_{\ell_1 \ell_3}$ to an sum over other other spherical waves, which is analogous to how crossing affects massive spinning particles (see \cite{Trueman:1964zzb, cohen-tannoudji_kinematical_1968} and more recently \cite{deRham:2017zjm}). 
},
\begin{align}
 \omega_1 = \frac{\omega_s}{2} \frac{1-\rho_s^2}{1- \rho_s \cos \theta_1} = \frac{\omega_s}{2} \left( 1 - \rho_s \cos \vartheta_1 \right) \; . 
 \label{eqn:psi_def}
\end{align}
The spherical wave expansion \eqref{eqn:sw_general} can then be written,
\begin{align}
&\mathcal{A}_s ( s, t , \omega_s, \omega_t, \omega_u )  \label{eqn:sw} \\
&= 64 \pi^2 \sum_{\substack{\ell_1 \ell_3 \\ m_1 m_3  }} Y^{m_1}_{\ell_1} ( \vartheta_1, \varphi_1 )  Y^{m_3*}_{\ell_3} ( \vartheta_3, \varphi_3 )  a^{m_1 m_3}_{\ell_1 \ell_3} ( \omega_s , \rho_s )
\nonumber
\end{align}
as described in \cite{Grall:2020tqc} \footnote{
Although note that \cite{Grall:2020tqc} uses the true angles $\theta_j$, whereas here we define the spherical waves \eqref{eqn:sw} using the eccentric angles $\vartheta_j$.
}, where $\varphi_j$ is the azimuthal angle of $\hat{\bfk}_j$ when $\bfk_s$ is aligned with the $z$-axis. We refer the reader there for the explicit $a^{m_1 m_3}_{\ell_1 \ell_3}$ from the EFT \eqref{eqn:EFT_action}. 

These angles are related to the analytic variables by,
\begin{align}
\omega_t &= - \frac{k_s}{2} \left( \cos \vartheta_1 - \cos \vartheta_3 \right) \;\; , \;\; t = - \frac{s}{2} \left(  1  - \cos \vartheta_{13}   \right)  \nonumber \\ 
\omega_u &= - \frac{k_s}{2} \left( \cos \vartheta_1 + \cos \vartheta_3 \right)  \; . 
\end{align}
where $\cos \vartheta_{13} = \cos \vartheta_1 \cos \vartheta_3 + \sin \vartheta_1 \sin \vartheta_3 \cos ( \varphi_{1} - \varphi_3 ) $ is the relative angle between $\bfk_1$ and $\bfk_3$. 
Each choice of $\{ \omega_s, \rho_s, \vartheta_1, \vartheta_3, \varphi_{1} - \varphi_3 \}$ corresponds to a unique $\{ s,t,\omega_s, \omega_t, \omega_u \}$ in the $s$-channel region.

~\\
{\bf Symmetries.}
The spherical wave coefficients in \eqref{eqn:sw} are constrained by various symmetries. For example, despite the centre-of-mass motion $\bfk_s$ breaking two of the rotational symmetries, there is one remaining rotation (rotations about $\bfk_s$) which guarantees that $\mathcal{A}_s$ depends only on the relative azimuthal angle $\varphi_1 - \varphi_3$. Furthermore, a spatial parity transformation corresponds to $\varphi_j \to \pi - \varphi_j $, and a time inversion corresponds to sending $\mathcal{A}_s  \to \mathcal{A}_{\bar s}$, which has the analogous spherical wave expansion with the roles of $\bfk_1 \bfk_2$ and $\bfk_3 \bfk_4$ exchanged.   
Finally, since we are scattering identical scalar fields, whether we use $\bfk_1$ and $\bfk_3$ or $\bfk_2$ and $\bfk_4$ to define the partial waves should make no difference, and consequently \eqref{eqn:sw} should be invariant under replacing $\cos \vartheta_1$ with $\cos \vartheta_2 = - \cos \vartheta_1$ and $\cos \vartheta_3$ with $\cos \vartheta_4 = - \cos \vartheta_3$. Altogether, these symmetries require,
\begin{align}
&\text{Rotational symmetry:} \;\; &a^{m_1 m_3}_{\ell_1 \ell_3}  &= 0 \;\;\; \text{unless} \;\; m_1 = m_3 \nonumber  \\ 
&\text{Bose symmetry:} \;\; &a^{m_1 m_3}_{\ell_1 \ell_3}   &= 0 \;\;\; \text{unless} \;\; \ell_1 + \ell_3 \;\; \text{even} \nonumber  \\ 
&\text{Parity:} \;\; &a^{m_1 m_3}_{\ell_1 \ell_3}   &= a^{-m_1 -m_3}_{\ell_1 \ell_3}  \nonumber  \\ 
&\text{Time Reversal:} \;\; &a^{m_1 m_3}_{\ell_1 \ell_3}  &= a^{m_3 m_1}_{\ell_3 \ell_1} 
\label{eqn:sw_symm}
\end{align}

~\\
{\bf Elastic Unitarity Bound.}
Since every term in \eqref{eqn:sw_unitarity} is positive, we can write an inequality involving only the $n=2$ term, 
\begin{align}
\text{Disc} \, a^{m_1 m_3}_{\ell_1 \ell_3} \geq \sum_{\substack{ J \\ m }} a_{\ell_1 J}^{m_1 m} a_{\ell_3 J}^{m_3 m *}
\label{eqn:unit_elastic}
\end{align}
This is now a non-linear bound which the $2 \to 2$ amplitude must satisfy as a consequence of unitarity.  
Note that for any complex unit vector $\hat{v}_{\ell}^m$, we can resolve the identity as $\delta_{J J'} = \hat{v}^{m*}_J \hat{v}_{J'}^{m'} + ...$ and write \eqref{eqn:unit_elastic} as,
\begin{align}
 \text{Im} \, \left[   \sum_{\substack{\ell_1 ,\ell_3 \\ m_1, m_3} } \hat{v}_{\ell_1}^{m_1} a_{\ell_1 \ell_3}^{m_1 m_3} \hat{v}_{\ell_3}^{m_3 *}   \right]  \geq  \Bigg|     \sum_{\substack{\ell_1 ,\ell_3 \\ m_1, m_3} } \hat{v}_{\ell_1}^{m_1} a_{\ell_1 \ell_3}^{m_1 m_3} \hat{v}_{\ell_3}^{m_3 *}   \Bigg|^2  \; .
 \label{eqn:unit_Imvav}
\end{align}
Since $| z| \geq \text{Im} \, z$ for any complex $z$, and the unit vector $\hat{v}_{\ell}^m$ can be written as a general vector $v_{\ell}^m$ divided by its norm, then \eqref{eqn:unit_Imvav} implies 
an upper bound on the spherical wave coefficients,
\begin{align}
\Bigg|     \sum_{\substack{\ell_1 ,\ell_3 \\ m_1, m_3} } v_{\ell_1}^{m_1} a_{\ell_1 \ell_3}^{m_1 m_3} v_{\ell_3}^{m_3 *}   \Bigg| \leq  \sum_{\substack{\ell \\ m} } \big| v_{\ell}^m \big|^2
\label{eqn:unit_vav}
\end{align}
for any complex vector, $v_{\ell}^m$. This is the analogue of the unitarity bound $| a_{\ell} | < 1$ in the Lorentz-invariant case.

~\\
{\bf Convergence and Boundedness.}
The partial wave expansion can also be used to derive the Froissart-Martin bound on the high energy growth of the amplitude in the absence of Lorentz boosts. 
This proceeds analogously to the Lorentz invariant proof, see for instance \cite{Gribov:2003nw}. 
Since the large $\ell$ behaviour of the spherical harmonics is $Y^{m}_{\ell} (\theta, \phi )  \sim e^{\ell | \text{Im} \, \theta |}$, convergence of the spherical wave expansion \eqref{eqn:sw} requires that the spherical waves fall off at large $\ell$ like,
\begin{align}
\lim_{\ell_1, \ell_3 \to \infty}  a^{m_1 m_3}_{\ell_1 \ell_3} \sim e^{ - \ell_1 \eta_1 - \ell_3 \eta_3}  \;\;  ,
\end{align}
for some $\eta_j ( \omega_s, \rho_s ) \geq | \text{Im} \, \vartheta_j |$. This allowed domain of the $\eta_j$ defines a region analogous to the Lehmann ellipse \cite{Lehmann:1958ita} in the Lorentz invariant case. Using the definition of $\vartheta_j$ above, at large $s$ we have, $\cos \vartheta_1 = 1 + z_+/s + ...$ and $\cos \vartheta_3 = 1 + z_-/s + ...$, where $z_{\pm} = 4 M ( 2 M ( \gamma - 1) \pm \omega_t) - t$ are held fixed \footnote{
Note that we are focussing on the large $\ell$ behaviour of the spherical wave expansion and neglecting the dependence on $m$. It is straightforward to retain the $m$ dependence (and use the large $s$ expansion of $\cos (\varphi_{1} - \varphi_3 )$) and this does not affect our final conclusion.
}. 
Consequently, convergence of the spherical wave expansion up to some fixed thresholds $z^{\rm th}_{\pm}$ requires, 
\begin{align}
\lim_{ \substack{  \ell_1, \ell_3 \to \infty \\ s \to \infty  }}  a^{m_1 m_3}_{\ell_1 \ell_3} \sim e^{ - ( \ell_1 \sqrt{z^{\rm th}_+}  +  \ell_3 \sqrt{z^{\rm th}_- } ) /  \sqrt{ s } }  \;\;  .
\end{align}
Further assuming polynomial boundedness (i.e. that $\text{Disc} \, \mathcal{A}_s$ does not grow faster than $s^{n}$ for some $n$), this means that the sums in $\ell_1$ and $\ell_3$ can be truncated at a maximum $\ell^*_\pm \sim \sqrt{s/z^{\rm th}_{\pm} } \log ( s^{n} )$ (since the neglected terms provide an exponentially small correction). 
This means that at large $s$ (with $t, M, \omega_t,  \gamma$ held fixed), 
\begin{align}
\lim_{s \to \infty} \mathcal{A}_s \sim 16 \pi \sum_{\ell_1 = 0}^{\ell^*_+ } \sum_{\ell_3 = 0}^{\ell^*_- } \sqrt{2 \ell_1 + 1} \sqrt{2 \ell_3 + 1} \; a^{00}_{\ell_1 \ell_3}
\end{align} 
where we have used that $Y^m_{\ell} = \delta^m_0 \sqrt{ 2 \ell + 1} / \sqrt{4 \pi }$ as $\vartheta \to 0$. From the unitarity bound \eqref{eqn:unit_vav} (with $v_{\ell}^0$ chosen as $\sqrt{2 \ell + 1}$ if $\ell \leq  \ell_* = \text{max} ( \ell^*_+, \ell^*_- )$ and $0$ otherwise), we have that $\lim_{s \to \infty} \mathcal{A}_s \lesssim \sum_{\ell = 0}^{\ell_*} (2 \ell_* + 1 ) \sim \ell_*^2$, and therefore,
\begin{align}
 \lim_{s \to \infty} \mathcal{A}_s  \lesssim s \log^2 s 
\end{align}  
when $t, M, \omega_t,  \gamma$ are fixed.
This coincides with the usual Froissart-Martin bound which constrains Lorentz-invariant amplitudes, i.e. the presence of a symmetry-breaking direction $n^\mu$ does not affect the very high energy (small-scale) behaviour of the amplitude, which remains bounded as if fully Poincar\'{e}-invariant.

~\\
{\bf Other Channels.}
Above we have focussed on the $s$-channel, in which $s > -t > 0$ and $| \omega_u \pm \omega_t | < \sqrt{\omega_s^2 - s}$. When we consider the analytically continued amplitude $\mathcal{A}_s (s, t , \omega_s, \omega_t, \omega_u)$ for values outside of this region, we can cross over to other channels. While \eqref{eqn:sw} can only be used in the $s$-channel region, there is an analogous partial wave expansion for each of the other five channels. For instance, for the $u$-channel, in which $u > -t > 0$ and $| \omega_s \pm \omega_t | < \sqrt{\omega_u^2 - u}$, we can write,
\begin{align}
&\mathcal{A}_s ( s, t , \omega_s, \omega_t, \omega_u ) 
= 
\mathcal{A}_u ( u, t , \omega_u, \omega_t, \omega_s )
  \label{eqn:sw_u} \\
&= 64 \pi^2 \sum_{\substack{\ell_1 \ell_3 \\ m_1 m_3  }} Y^{m_1}_{\ell_1} ( \vartheta_1^u , \varphi_1^u )  Y^{m_3*}_{\ell_3} ( \vartheta_3^u , \varphi_3^u  )  b^{m_1 m_3}_{\ell_1 \ell_3} ( \omega_s , \rho_s )
\nonumber
\end{align} 
using the crossing \eqref{eqn:crossing}, where $\vartheta_j^u$ are the eccentric angles of $\bfk_j$ relative to $\bfk_u = \bfk_1 - \bfk_4$, $\varphi_j^u$ are the corresponding azimuthal angles, and $b^{m_1 m_3}_{\ell_1 \ell_3}$ are the spherical waves for the $u$-channel process (and are equivalent to $a^{m_1 m_3}_{\ell_1 \ell_3}$ when scattering identical scalars). There is a spherical wave expansion for each of the six channels, and to which the preceding arguments apply.

~\\
{\bf Examples.}
In order to illustrate the properties of the spherical wave coefficients described above, we provide here the $a^{m_1 m_3}_{\ell_1 \ell_3}$ for two simple interaction terms. 
We concentrate on quartic interactions, for which there are finitely many non-zero spherical wave coefficients \footnote{
In contrast, cubic interactions---which generate exchange contributions to $2\to2$ scattering---can generate infinitely many partial wave coefficients due to the $t$- and $u$-channel poles. For further details on how to compute these we refer the reader to \cite{Grall:2020tqc}.
}, and in particular a $\beta_1 \dot\pi^4$ interaction (which appears in \eqref{eqn:EFT_action} at leading order) as well as a $ \beta_t\,\dot \pi \ddot\pi^3$ interaction (which is the first non-trivial, \emph{non} time reversal invariant interaction at lowest order in derivatives). 
The amplitudes corresponding to each interaction are:
\begin{align}
    \A[\beta_1]&=4! \beta_1 \prod_{i=1}^4 \omega_i   ~ , \;\; 
    &\A[\beta_t]&=3 \omega_s \omega_t \omega_u \prod_{i=1}^4 \omega_i \,,
\end{align}
and the corresponding spherical wave coefficients are:
\begin{align}
    a_{00}^{00}[\beta_1]&=\frac{\omega_s^4}{\pi}\frac{(3-\rho_s^2)^2}{96}\,,\\
    a_{02}^{00}[\beta_1]&= a_{20}^{00}[\beta_1]=\frac{\omega_s^4}{\pi}\frac{\rho_s^2(-3+\rho_s^2)}{48\sqrt{5}}\,,\\
    a_{22}^{00}[\beta_1]&=\frac{\omega_s^4}{\pi}\frac{\rho_s^4}{120}\,,
\end{align}
and,
\begin{align}
    a_{02}^{00}[\beta_t]&=-a_{20}^{00}[\beta_t]=\frac{\omega_s^7}{\pi}\frac{\rho_s^2(-3\rho_s^4+30\rho_s^2-35)}{17920\sqrt{5}}\,,\\
    a_{04}^{00}[\beta_t]&=-a_{40}^{00}[\beta_t]=\frac{\omega_s^7}{\pi}\frac{\rho_s^4(3-\rho_s^2)}{13440}\,,\\
    a_{24}^{00}[\beta_t]&=-a_{42}^{00}[\beta_t]=-\frac{\omega_s^7}{\pi}\frac{\rho_s^6}{6720\sqrt{5}}\, ,
\end{align}
where $\rho_s^2 = 1 - s/\omega_s^2$. 
Notice that these respect the relevant symmetries \eqref{eqn:sw_symm} we mentioned earlier. 
In particular the spherical waves for the $\beta_t$ interaction, which is odd under time reversal, are \emph{anti}-symmetric under $\ell_1\leftrightarrow\ell_3$ (unlike the $\beta_1$ spherical waves which are symmetric, since the $\dot \pi^4$ interaction is even under time reversal).


\section{More Positivity Bounds}
\label{app:more_pos_bounds}

\noindent In the main text we have derived the forward limit positivity bounds \eqref{eqn:pos_fwd}. 
In this Appendix, we describe a strategy which can be used to generate a further infinite family of bounds on the $t$ and $\omega_1 \omega_3$ derivatives of the EFT amplitude. 
 
~\\
{\bf Beyond the Forward Limit.}
In the Lorentz-invariant case, positivity of $\text{Disc} \, \A_s(s,t)$ can be extended beyond the forward limit to include any number of $t$ derivatives by exploiting properties of the partial wave expansion \cite{Nicolis:2009qm, deRham:2017avq}. 
To go beyond the forward limit when boosts are broken, the partial wave expansion must be replaced with the spherical wave expansion recently derived in \cite{Grall:2020tqc}, and we have proven in Appendix~\ref{sec:sw} that the optical theorem \eqref{eqn:optical} implies, 
\begin{align}
\left( \frac{\partial}{\partial t}  \right)^i \left(  \frac{\partial^2}{\partial \omega_1 \partial \omega_3} \right)^j \text{Disc} \, \mathcal{A}_s 
\bigg|_{ \substack{t=0  \\ \omega_t = 0}} \geq 0  \; , 
\label{eqn:unit_pos}
\end{align}
for any integer $i$ and $j$ and for any forward-limit $s$-channel kinematics. Note that the $\omega_1 \omega_3$ derivative can be written as $\partial_{\omega_u}^2 - \partial_{\omega_t}^2$ and so these bounds correspond to going beyond the function evaluated at $t=0$ and $\omega_t =0$ to include arbitrary derivatives $(\partial_t)^i$ and $(\partial_{\omega_t})^{2j}$ at these points.

~\\
{\bf Dispersion Relation.}
The goal is then to use a dispersion relation to connect the more general positivity properties \eqref{eqn:unit_pos} of the UV theory with the low-energy EFT amplitude.
As discussed in the main text, this requires specifying which variables will be held fixed as one analytically continues into the complex $s$ plane. Rather than the boosted Breit coordinates $\{ \gamma, M \}$ from equation \eqref{eqn:BoostedBreit}, in this Appendix we will use the variables $\{ E_s, E_u \}$, defined by, 
\begin{align}
 \omega_s = \frac{s + 4M^2}{4 M} + E_s \;\;, \;\;  \omega_u = \frac{u + 4M^2}{4 M} + E_u
\label{eqn:f_eg}
\end{align}
where $M$ is a fixed mass scale. Once the amplitude is written in terms of these variables, which we denote by $
\hat{ \mathcal{A}}_s (s,t, E_s, \omega_t, E_u )$ ($= \mathcal{A}_s (s,t, \omega_s , \omega_t, \omega_u)$ with $\omega_s$ and $\omega_u$ replaced by \eqref{eqn:f_eg}), crossing \eqref{eqn:crossing} becomes,
\begin{align}
 \hat{A}_s ( s, t , E_s , \omega_t, E_u ) = \hat{A}_u ( u, t, E_u, \omega_t, E_u) \; ,
\end{align}
and so the kernel which appears in the dispersion relation \eqref{eqn:dispersion} is now given by,
\begin{align}
P_n (\mu, s)  =  \frac{ \text{Im} \, \hat{\mathcal{A}}_s (  \mu , t , E_s, \omega_t, E_u )   }{(\mu-s)^{n+1} } -  \frac{ \text{Im} \, \hat{\mathcal{A}}_u (  \mu , t , E_u , \omega_t, E_s )   }{(u - \mu )^{n+1} }   .
\label{eqn:Pn_EsEu}
\end{align}
Despite not being manifestly crossing symmetric, it is possible to keep $\{ E_s, E_u \}$ fixed so that both branch cuts have physical kinematics---this corresponds to satisfying the condition \footnote{
When $E_s$ or $E_u$ is negative, \eqref{eqn:EsEu_cond} must be supplemented by $\sqrt{s} > 2M + 2 \sqrt{-E_s M}$ and $\sqrt{s} > 2M + 2 \sqrt{-E_u M}$, otherwise there is a interval of $s$ near $4M^2$ for which $| \bfk_s | = \sqrt{\omega_s^2 - s}$ becomes imaginary. 
},
\begin{align}
s  \; \frac{ E_s + E_u }{ | E_s - E_u | }  >  2 M \left( 2 M + E_s + E_u  \right)  \; . 
\label{eqn:EsEu_cond}
\end{align}
This condition is always satisfied for sufficiently large $s$, and in particular when $E_s = E_u$ it is always satisfied.
The reason for the change of notation is purely pedagogical---since $\{ E_s, E_u \}$ are not manifestly crossing symmetric, they will better highlight some of the challenges in going beyond the forward limit (and how to overcome them).

~\\
{\bf $\partial_t$ Bound.}
In the Lorentz-invariant case, taking $t$-derivatives of the dispersion relation was shown in \cite{deRham:2017avq} to produce a second infinite series of bounds. In that case, although $\partial_t \, \text{Im} \, \mathcal{A}_s$ is positive, the $\partial_t$ also acts on the $1/(\mu-u)^{n+1}$ of the crossed branch cut, and this gives a negative contribution to the dispersion relation for $\partial_t \partial_s^{2n} \mathcal{A}_s$. The remedy is to compensate for this negative term by adding a sufficiently positive lower derivative, which produces bounds such as,
\begin{align}
\left( \partial_t + \frac{2n+1}{s_b}  \right) \partial_s^{2n} \mathcal{A}_s   \geq 0 \; , 
\label{eqn:pos_dt_LI}
\end{align}
in the Lorentz-invariant case \footnote{
Note we have not set $s=2m^2-t/2$ and exploited $s$-$u$ crossing to combine the $\text{Im} \, \tilde{ \mathcal{A}}_s$ and $\text{Im}\, \tilde{\mathcal{A}}_u$ appearing in \eqref{eqn:Pn}, and hence there is a factor of 2 difference between \eqref{eqn:pos_dt_LI} and the bound of \cite{deRham:2017avq}.
}.

In our case, we no longer have Lorentz boosts and the dispersion relation is now given by \eqref{eqn:dispersion} and \eqref{eqn:Pn_EsEu}. Once $\omega_s$ and $\omega_u$ have been replaced by explicit functions of $t$, a simple $t$ derivative of $\text{Disc}\, \hat{\mathcal{A}}_s$ is no longer positive, since
\begin{align}
 \partial_t \hat{\mathcal{A}}_s  = \left( \partial_t  - \frac{1}{4M} \partial_{\omega_u} \right)
 \mathcal{A}_s
\end{align}
and $\partial_{\omega_u}$ of $\text{Disc} \, \mathcal{A}_s$ is not sign definite.
Since the dispersion relation involves both $\hat{\mathcal{A}}_s$ and $\hat{\mathcal{A}}_u$, it is \emph{not} possible to compensate for this $ \partial_{\omega_u}$ simultaneously on both cuts. 
However, the \emph{integral} of this term with respect to $E_u$ is positive. 
In fact, if we define,
\begin{align}
\mathcal{I} [ \hat{\mathcal{A}}_s ] 
 = \int_0^{E_s} d E_s' \int_0^{E_u} d E_u' \; \hat{\mathcal{A}}_s (s,t, E_s', \omega_t, E_u' )
\label{eqn:IA}
\end{align}
then we have that,
\begin{align}
 \partial_{E_s} \text{Disc} \, \mathcal{I} [ \hat{\mathcal{A}}_s ] \geq 0  \;\;\;\; \text{and} \;\;\;\;  \partial_{E_u} \text{Disc} \, \mathcal{I} [ \hat{\mathcal{A}}_s ] \geq 0 
 \label{eqn:dEsIA}
\end{align}
when $\omega_t^2 = t = 0$, which follows from \eqref{eqn:unit_pos} with $i=j=0$, and also that,
\begin{align}
\left( \partial_t  + \frac{1}{4M}  \partial_{E_u}  \right) \text{Disc} \, \mathcal{I} [ \hat{\mathcal{A}}_s ] \geq 0  \; , 
\label{eqn:dtIA}
\end{align}
when $\omega_t^2 = t =0$, thanks to \eqref{eqn:unit_pos} with $i=1$, $j=0$. 
Integrating the dispersion relation with respect to $E_s$ and $E_u$ therefore provides the analogue of the Lorentz-invariant bound \eqref{eqn:pos_dt_LI} when boosts are broken,
\begin{align}
\noindent \left( \partial_t + \tfrac{2n+1}{s_b}  + \tfrac{1}{4 M}  \partial_{E_s}  +  \tfrac{1}{4M}  \partial_{E_u}  \right) \mathcal{I} [  \hat{\mathcal{A}}_s^{(2N)} ] \big|_{ \substack{ t=0 \\ \omega_t = 0 } } \geq 0  \hfill
\label{eqn:pos_dt}
\end{align}
where again $N \geq 1$ and $E_s, E_u$ must be physical on both cuts \eqref{eqn:EsEu_cond}. 
To recap, $\mathcal{I} [  \tilde{\mathcal{A}}_s^{(2n)} ]$ is the EFT scattering amplitude $\mathcal{A}_s (s,t, \omega_s, \omega_t, \omega_u )$ with $\omega_s$ and $\omega_u$ replaced by $E_s$ and $E_u$ as in \eqref{eqn:f_eg} and then a portion of the branch cut (up to a scale $s_b$ at which the EFT remains reliable) subtracted as in \eqref{eqn:improv} and finally integrated over $E_s$ and $E_u$ as in \eqref{eqn:IA}. This object is straightforward to compute in practice, and we have demonstrated that the new positivity bound \eqref{eqn:pos_dt} may be used to diagnose whether a Lorentz invariant, unitarity, causal, local UV completion of the EFT is possible. 

While it is straightforward to substitute the dispersion relation for $\hat{\A}_s$ into \eqref{eqn:pos_dt}  and use \eqref{eqn:dEsIA} and \eqref{eqn:dtIA} to confirm that it is indeed positive, let us now re-derive \eqref{eqn:pos_dt} using a more systematic approach which can be readily generalised to higher derivatives. 

~\\
{\bf From Positivity to Monotonicity.}
To systematically go to higher derivatives/integrals of $\hat{\mathcal{A}}_s$, it is helpful to introduce the indefinite integrals, $\hat{\mathcal{A}}_{ab} (E_s, E_u)$, which obey,
\begin{align}
\partial_{E_s}^a \partial_{E_u}^b  \hat{\mathcal{A}}_{ab} ( E_s , E_u ) =  \hat{\A}_s ( s, t , E_s, \omega_t, E_u ) \; . 
\end{align}
Since the original $\text{Disc} \, \hat{A}_s$ was \emph{positive} for all $E_s$ and $E_u$, then the first integrals $\text{Disc} \, \hat{\A}_{10}$ and $\text{Disc} \, \hat{\A}_{01}$ are \emph{monotonic} in $E_s$ and $E_u$ respectively. This means that the combinations,
\begin{align}
 I_{10} ( E_s , \delta_s ; E_u ) &= \hat{\A}_{10} (E_s + \delta_{s} , E_u) - \hat{\A}_{10} (E_s + \delta_{s} , E_u)    \nonumber  \\
 I_{01} ( E_s ;  E_u , \delta_u ) &= \hat{\A}_{01} (E_s , E_u + \delta_{u} ) - \hat{\A}_{01} (E_s , E_u+ \delta_{u} )     
\end{align}
have $\text{Disc} \, I_{10} \geq 0$ and $\text{Disc} \, I_{01} \geq 0$  for all $\delta_s \geq 0$. The double integral $\text{Disc}\, \hat{\A}_{11} (E_s, E_u)$ is ``doubly monotonic'', in that the following quantity,
\begin{align}
I_{11} ( E_s, \delta_s ; E_u , \delta_u ) &= \hat{\A}_{11} (E_s + \delta_{s} , E_u + \delta_u) + \hat{\A}_{11} (E_s  , E_u)   \nonumber \\
&-  \hat{\A}_{11} (E_s + \delta_s  , E_u ) - \hat{\A}_{11} ( E_s  , E_u + \delta_u )
\end{align}
has $\text{Disc} \, I_{11} \geq 0$ for all $\delta_s \geq 0$ and $\delta_u \geq 0$. Note that $I_{11} (0, E_s ; 0 , E_u)$ is nothing more than the $\mathcal{I} [ \mathcal{A}_s ]$ defined earlier. 
This is useful because it makes transparent the action of each derivative,
\begin{align}
&\partial_{\delta_s} I_{11} (E_s, \delta_s ; E_u, \delta_u ) = I_{01} ( E_s + \delta_s ; E_u , \delta_u  ) \nonumber  \\[5pt]
&\partial_{E_s} I_{11} (E_s, \delta_s ; E_u, \delta_u ) \nonumber \\
&= I_{01} ( E_s + \delta_s ; E_u , \delta_u  ) - I_{01} ( E_s ; E_u , \delta_u  ) \; .
\label{eqn:dI11}
\end{align}

Similarly, since $\left( \partial_t + \tfrac{1}{4M} \partial_{E_u} \right) \text{Disc} \, \hat{A}_s (s,t,E_s,\omega_t, E_u )$ is positive for all $E_s$ and $E_u$, then when we take two integrals the resulting $\left( \partial_t + \tfrac{1}{4M} \partial_{E_u} \right) \text{Disc} \, \hat{A}_{11} ( E_s, E_u )$ is ``doubly monotonic'', so we have that,
\begin{align}
\left( \partial_t + \frac{1}{4M} \partial_{E_u}  \right) \text{Disc}\, I_{11} (E_s, \delta_s ; E_u , \delta_u ) \geq 0 
\end{align}
The issue that arises when embedding this into a dispersion relation such as,
\begin{align}
&\frac{1}{2N!} \partial_s^{2N} I_{11} (E_s, \delta_s ; E_u, \delta_u )  \nonumber \\
&= \int \frac{d\mu}{\pi} \left[  \frac{\text{Im} \, I_{11} (E_s, \delta_s ; E_u , \delta_u) }{ (\mu - s )^{2N+1} }  +
\frac{\text{Im} \, I_{11} ( E_u , \delta_u ; E_s, \delta_s  ) }{ (\mu - u )^{2N+1} }
   \right]
   \label{eqn:Idisp}
\end{align}
is that the $E_u$ derivative does not affect both branch cuts in the same way. 
In particular, from \eqref{eqn:dI11} we see that $\partial_{E_u}$ acting on the $\text{Disc} \, I_{11} (E_u, \delta_u ' E_s, \delta_s )$ is \emph{not} sign definite.
In order to have positive $t$ derivatives on both cuts, we need to use a crossing symmetric derivative $\partial_{E_s} + \partial_{E_u}$, and in order for the $\partial_{E_s}$ ($\partial_{E_u}$) derivative on the right (left) cut to be positive we need to compensate for the negative term appearing in \eqref{eqn:dI11}. That is to say, it is only the combination,
\begin{align}
 \left( \partial_t + \frac{1}{4M} \partial_{E_s} + \frac{1}{4M} \partial_{E_u}   \right) I_{11} + 
 I_{01} + I_{10} \geq 0 \; , 
\end{align}
which has positive $\text{Disc}$ on both cuts (i.e. for both arguments $(E_s, \delta_s;E_u,\delta_u)$ and $(E_u, \delta_u; E_s, \delta_s)$). This can be written more succinctly as,
\begin{align}
 \left( \partial_t + \frac{1}{4M} \partial_{\delta} \right) \text{Disc} \, I_{11} \geq 0 \; . 
\end{align}
where $\partial_\delta = \partial_{\delta_s} + \partial_{\delta_u}$. 

Finally, defining $D_t = \partial_t + \frac{2n+1}{s_b}$ just like in the Lorentz-invariant case \eqref{eqn:pos_dt_LI}, so that,
\begin{align}
 D_t \left[ \frac{ \text{Im} \, \hat{\A}_s (\mu) }{ (\mu - u)^{2n+1} } \right] \geq  \frac{ \partial_t \text{Im} \, \hat{\A}_s (\mu) }{ (\mu - u)^{2n+1} }
 \label{eqn:Dt}
\end{align}
for all $\mu \geq s_b$ when $u = 0$, then we arrive at the first $t$ derivative bound \eqref{eqn:pos_dt}, 
\begin{align}
 \left(  D_t  + \frac{1}{4M} \partial_{\delta} \right) I_{11}^{(2N)} ( E_s, \delta_s ; E_u, \delta_u ) \geq 0 \; .
\end{align}
where we have evaluated the amplitude at $s=t=\omega_t^2 =0$. This bound constrains $\partial_t \partial_s^{2N}$ of the EFT amplitude.

The positivity bounds \eqref{eqn:pos_fwd} and \eqref{eqn:pos_dt} follow from unitarity \eqref{eqn:unit_pos} in the UV with $j=0$ and $i=0$  or $i=1$. The construction we have described above generalises straightforwardly to any $i$ or $j$, and in general requires taking further energy integrals of the amplitude. 
We will now illustrate this by deriving positivity bounds from \eqref{eqn:unit_pos} with $i=2, j=0$ and $i=0,j=1$. 

~\\
{\bf From Monotonicity to Convexity.}
Taking two $E_s$ derivatives, the indefinite integral $\tilde{\A}_{20} (E_s, E_u)$ is now a \emph{convex} function,  and so $\text{Disc} \, \hat{\A}_{20}$ is only positive in the combination, 
\begin{align}
I_{20} (E_s, \delta_{s1} , \delta_{s2} ; E_u ) &= \hat{\A}_{11} (E_s + \delta_{s} , E_u + \delta_u) + \hat{\A}_{11} (E_s  , E_u)   \nonumber \\
&-  \hat{\A}_{11} (E_s + \delta_s  , E_u ) - \hat{\A}_{11} ( E_s  , E_u + \delta_u ) \; ,
\end{align}
and similarly $I_{02} (E_s ; E_u, \delta_{u1} , \delta_{u2} )$ is the analogous convex combination of $\hat{\mathcal{A}}_{02} (E_s, E_u)$. Taking two $E_s$ and one $E_u$ integral, $\hat{\A}_{21} (E_s, E_u)$ is ``convex-monotonic'', and so we must use the combination,
\begin{align}
&I_{21} ( E_s, \delta_{s1}, \delta_{s2}; E_u , \delta_u ) = \nonumber \\
&\int_{ E_s }^{ E_s + \delta_{s2} } d E_s'  \int_{ E_s'}^{ E_s'+\delta_{s1}} d E_s \; \int_{ E_u }^{ E_u + \delta_{u} } d E_u'   \tilde{A}_s ( s,t,E_s'' ,\omega_t, E_u' )
\end{align}
which is a linear combination of $\hat{\A}_{21}$ with 8 different arguments, in order to use $\text{Disc} I_{21} \geq 0$. 
Finally, taking two integrals of $E_s$ and $E_u$, $\hat\A_{22} (E_s, E_u)$ is ``doubly convex'', and the appropriate combination to take is,
\begin{align}
&I_{22} ( E_s , \delta_{s1}, \delta_{s2} ;  E_u , \delta_{u1}, \delta_{u2} )   \nonumber \\
&= \int_{ \bar{E}_s }^{\bar{E}_s + \delta_{s2} } d \bar{E}_s'  \int_{\bar{E}_s'}^{\bar{E}_s'+\delta_{s1}} d E_s \; \int_{ \bar{E}_s }^{\bar{E}_u + \delta_{u2} } d \bar{E}_u'  \int_{\bar{E}_u'}^{\bar{E}_u'+\delta_{u1}} d E_u \; \tilde{A}_s  \; . 
\end{align}
which can be written as a sum over $\hat{\A}_{22}$ with 16 different arguments. Again, these $I_{ab}$ combinations are convenient because they have positive discontinuities and their derivatives are related straightforwardly, e.g.
\begin{align}
& \left( \partial_{\delta_{u1}} - \partial_{E_u} \right) \partial_{\delta_{s2}} I_{22} ( E_s , \delta_{s1}, \delta_{s2} ;  E_u , \delta_{u1}, \delta_{u2} )    \nonumber \\
 &\qquad\qquad\qquad\qquad= I_{11} \left( E_s + \delta_{s2} , \delta_{s1} ; E_u , \delta_{u2} \right) 
\end{align} 
These are the natural building blocks with which to construct the $\partial_t^2$ and $\partial^2/\partial \omega_1 \partial \omega_3$ positivity bounds.

~\\
{\bf $\partial_t^2$ Bound.}
The positivity property \eqref{eqn:unit_pos} with $i=2$ and $j=0$ corresponds to,
\begin{align}
\left(  \partial_t + \frac{1}{4M} \partial_{E_u}  \right)^2 \text{Disc} \, I_{22} \geq 0 \; ,
\label{eqn:ddtI22}
\end{align}
but as with the first $t$-derivative this second derivative cannot be applied directly to a dispersion relation like \eqref{eqn:Idisp} for $I_{22}$ since it is not symmetric in $E_s \leftrightarrow E_u$. One can instead begin with the na\"{i}ve symmetrisation of \eqref{eqn:ddtI22}, $(\partial_t + \tfrac{1}{4M} \partial_{E_s} + \tfrac{1}{4M} \partial_{E_u}) I_{22}$, and then add lower order $I_{12}, I_{21}$, etc. in order to compensate for the negative terms which are generated on the left- and right-hand cuts.
One arrives at the following combination,
\begin{align}
 \left( \partial_t + \frac{1}{4M} \partial_{\delta_1} \right)  \left( \partial_t + \frac{1}{4M} \partial_{\delta_2} \right) \text{Disc} \, I_{22} \geq 0 \; . 
\end{align}
which is positive on both cuts (i.e. for both arguments $( E_s , \delta_{s1}, \delta_{s2} ;  E_u , \delta_{u1}, \delta_{u2} )$ and $(  E_u , \delta_{u1}, \delta_{u2} ; E_s , \delta_{s1}, \delta_{s2} ) $), where $\partial_{\delta_1} = \partial_{ \delta_{s1} } + \partial_{ \delta_{u1} }$.

To convert this into a constraint on the EFT amplitude, one must again promote the $\partial_t$ derivatives as in \eqref{eqn:Dt}.
This produces the positivity bound, 
\begin{align}
 \left( D_t + \frac{1}{4M} \partial_{\delta_1} \right)  \left( D_t + \frac{1}{4M} \partial_{\delta_2} \right) I_{22}^{(2N)} \geq 0 \; . 
\end{align}
which constrains $\partial_t^2 \partial_s^{2N}$ of the EFT amplitude.

~\\
{\bf $\partial^2/\partial \omega_1 \omega_3$ Bound.}
The optical theorem \eqref{eqn:unit_pos} also guarantees that $\partial_{\omega_u}^2 - \partial_{\omega_t}^2 = \partial^2/\partial\omega_1 \partial\omega_3$ acting on $\text{Disc} \, \A_s$ is positive. Proceeding analogously to the above, this can be written as the positivity bound,
\begin{align}
\left[ \partial_{E_s}^2 + \partial_{E_u}^2 - \partial_{\omega_t}^2 - ( \partial_{\delta_{s1}} - \partial_{\delta_{s2}}  )^2 - (  \partial_{\delta_{u1}} - \partial_{\delta_{u2}}  )^2 \right] I_{22}^{(2N)}  \geq 0 
\end{align}
on the EFT amplitude.

\bibliography{multifield_positivity}

\end{document}